\documentclass[12pt,preprint]{aastex}

\def\earth{\oplus}
\usepackage{epsfig}
\begin{document}
\title{Atmospheric Mass Loss During Planet Formation: The Importance of Planetesimal Impacts} \shortauthors{Schlichting et al.}
\shorttitle{Atmospheric Mass Loss} \author{Hilke E. Schlichting\altaffilmark{1}, Re'em Sari\altaffilmark{2} and Almog Yalinewich\altaffilmark{2}
}\altaffiltext{1} {Massachusetts Institute of Technology, 77 Massachusetts Avenue, Cambridge, MA 02139-4307, USA}\altaffiltext{2}{Racah Institute of Physics, Hebrew University, Jerusalem 91904, Israel} \email{hilke@mit.edu}

\begin{abstract}
Quantifying the atmospheric mass loss during planet formation is crucial for understanding the origin and evolution of planetary atmospheres. We examine the contributions to atmospheric loss from both giant impacts and planetesimal accretion. Giant impacts cause global motion of the ground. Using analytic self-similar solutions and full numerical integrations we find (for isothermal atmospheres with adiabatic index $\gamma=5/3$) that the local atmospheric mass loss fraction for ground velocities $v_g \lesssim 0.25 v_{esc}$ is given by $\chi_{loss}=(1.71 v_g/v_{esc})^{4.9}$, where $v_{esc}$ is the escape velocity from the target. Yet, the global atmospheric mass loss is a weaker function of the impactor velocity $v_{Imp}$ and mass $m_{Imp}$ and given by $X_{loss} \simeq 0.4x+1.4x^2-0.8x^3$ (isothermal atmosphere) and  $X_{loss} \simeq 0.4x+1.8x^2-1.2x^3$ (adiabatic atmosphere), where $x=(v_{Imp}m/v_{esc}M)$. Atmospheric mass loss due to planetesimal impacts proceeds in two different regimes: 1) Large enough impactors
$m \gtrsim \sqrt{2} \rho_0 (\pi h R)^{3/2}$ (25~km for the current Earth), are able to eject all the atmosphere above the tangent plane of the impact site, which is $h/2R$ of the whole atmosphere, where $h$, $R$ and $\rho_0$ are the atmospheric scale height, radius of the target, and its atmospheric density at the ground. 2) Smaller impactors, but above $m>4 \pi \rho_0 h^3$ (1~km for the current Earth) are only able to eject a fraction of the atmospheric mass above the tangent plane. We find that the most efficient impactors (per unit impactor mass) for atmospheric loss are planetesimals just above that lower limit (2~km for the current Earth). For impactor flux size distributions parametrized by a single power law, $N(>r) \propto r^{-q+1}$, with differential power law index $q$, we find that  for $1<q<3$ the atmospheric mass loss proceeds in regime 1) whereas for $q>3$ the mass loss is dominated by regime 2). Impactors with $m \lesssim 4 \pi \rho_0 h^3$ are not able to eject any atmosphere. Despite being bombarded by the same planetesimal population, we find that the current differences in Earth's and Venus' atmospheric masses can be explained by modest differences in their initial atmospheric masses and that the current atmosphere of the Earth could have resulted from an equilibrium between atmospheric erosion and volatile delivery to the atmosphere from planetesimal impacts. We conclude that planetesimal impacts are likely to have played a major role in atmospheric mass loss over the formation history of the terrestrial planets. 
\end{abstract}


\section{INTRODUCTION}
Terrestrial planet formation is generally thought to have proceeded in two main stages: The first consists of the accretion of planetesimals, which leads to the formation of several dozens of roughly Mars-sized planetary embryos  \citep[e.g.][]{IM93,WSDMO97}, and the second stage consists of a series of giant impacts between these embryos that merge to form the Earth and other terrestrial planets  \citep[e.g.][]{ACL99,C01}. Understanding how much of the planets' primordial atmosphere is retained during the giant impact phase is crucial for understanding the origin and evolution of planetary atmospheres. In addition, a planet's or proptoplanet's atmosphere cannot only be lost due to a collision with a comparably sized body in  a giant impact, but also due to much smaller impacts by planetesimals. During planet formation giant impacts begin when the planetesimals are no longer able to efficiently damp the eccentricities of the growing protoplanets. Order of magnitude estimates that balance the stirring rates of the protoplanets with the damping rates due to dynamical friction by the planetesimal population and numerical simulations find that giant impacts set in when the total mass in protoplanets is comparable to the mass in planetesimals \citep{GLS042,KB06}. Therefore about 50\% of the total mass still resides in planetesimals when giant impacts begin and planetesimal accretion continues throughout the giant impact phase.  Furthermore, geochemical evidence from highly siderophile element (HSE) abundance patterns inferred for the terrestrial planets and the Moon suggest that a total of about $0.01~M_{\earth}$ of chondritic material was delivered as `late veneer' by planetesimals to the terrestrial planets after the end of giant impacts \citep{W99,WH04,W09}. This suggests that planetesimal accretion did not only proceed throughout the giant impacts stage by continued beyond. Therefore, in order to understand the origin and evolution of the terrestrial planets' atmospheres one needs to examine the contribution to atmospheric loss from both the giant impacts and from planetesimal accretion.

Depending on impactor sizes, impact velocities and impact angles, volatiles may be added to or removed from growing planetary embryos by impacts of other planetary embryos and smaller planetesimals. The survival of primordial atmospheres through the stage of giant impacts during terrestrial planet formation has been examined by \citet{GA03} and \citet{GA05}. These works numerically integrate the hydrodynamic equations of motion of the planetary atmosphere to determine the amount of atmospheric loss for various ground velocities. 
In contrast to giant impacts, smaller impactors cannot eject the planet's atmosphere globally but are limited to, at best, ejecting all the atmosphere above the tangent plane of the impact site. 
Some of the first calculations of impact induced atmospheric erosion were performed using the \citet{ZR67} solution for the expansion of a vapor plume and momentum balance between the expanding gas and the mass of the overlying atmosphere \citep[e.g.][]{MV89,VM90,A93}. The results of these calculations were used to investigate the evolution of planetary atmospheres as a result of planetesimal impacts \citep[e.g.][]{Z90,ZP92}. \citet{N99} investigated by analytical and computational means the effect of $\sim 10$~km impactors on terrestrial atmospheres using an analytical model based on the solutions of \citet{K60}. Atmospheric erosion calculations were extended further by, for example, \citet{S07} and \citet{Shu09} who investigated numerically atmospheric loss and replenishment and the role of oblique impacts, respectively.
In the work presented here, we use order of magnitude estimates and numerical simulations to calculate the atmospheric mass loss over the entire range of impactor sizes, spanning impacts too small to eject significant amounts of atmosphere to planetary-embryo scale giant impacts. Our results demonstrate that the most efficient impactors (per impactor mass) for atmospheric loss are small planetesimals which, for the current atmosphere of the Earth, are only about 2~km in radius.  We show that these small planetesimal impacts could have potentially totally dominated the atmospheric mass loss over Earth's history and during planet formation in general.

Our paper is structured as follows: In section 2 we use analytic self-similar solutions and full numerical integrations to calculate the amount of atmosphere lost during giant impacts for an isothermal and adiabatic atmosphere. We analytically calculate the atmospheric mass loss due to planetesimal impacts in section 3. In section 4, we compare and contrast the atmospheric mass loss due to giant impacts and planetesimal accretion and show that planetesimal impacts likely played a more important role for atmospheric loss of terrestrial planets than giant impacts. We discuss the implications of our results for terrestrial planet formation and compare our findings with recent geochemical constraints on atmospheric loss and the origin of Earth's atmosphere in section 5. Discussion and conclusions follow in section 6.

\section{Atmospheric Mass Loss Due to Giant Impacts}\label{s1}
When an impact occurs the planet's atmosphere can be lost in two distinct ways: First, the expansion of plumes generated at the impact site can expel the atmosphere locally but not globally. Atmospheric loss is therefore limited to at best $h/(2R)$ of the total atmosphere, where $h$ is the atmospheric scale height and $R$ the planetary radius (see section \ref{s2} for details). Second, giant impacts create a strong shock that propagates through the planetary interior causing a global ground motion of the proto-planet. This ground motion in turn launches a strong shock into the planetary atmosphere, which can lead to loss of a significant fraction of or even the entire atmosphere. 

\begin{figure}[htp]
   \centering
   \includegraphics[width=0.9\textwidth]{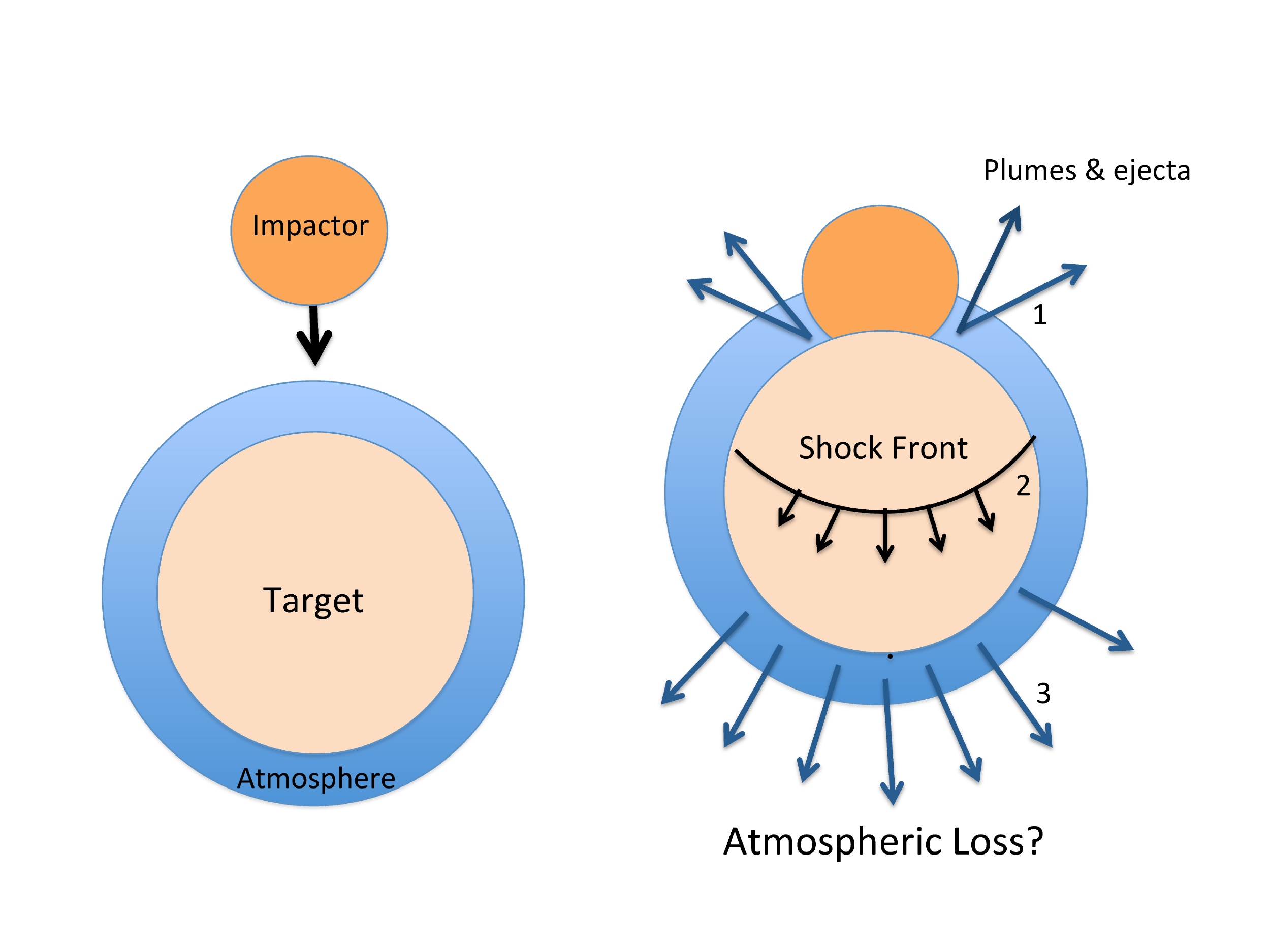}
\caption{Illustration of a giant impact. 1) The giant impact ejects atmosphere and ejecta close to the impact point and launches a strong shock. 2) The shock front propagates through the target causing a global ground motion. 3) This ground motion in turn launches a strong shock into the planetary atmosphere, which can lead to loss of a significant fraction of or even the entire atmosphere.}
\label{fig1}
\end{figure} 

It was realized several decades ago that self-similar solutions provide an excellent description for a shock propagating in adiabatic and isothermal atmospheres \citep[e.g.][]{R64,GH66}. Here, we take advantage of these self-similar solutions and use them together with full numerical integrations to calculate the atmospheric mass loss due to giant impacts. 

\subsection{Self-Similar Solutions to the Hydrodynamic Equations for an Isothermal Atmosphere}
Terrestrial planet's atmospheres, like the Earth's, are to first order isothermal, giving rise to an exponential density profile. We therefore solve the hydrodynamic equations for a shock propagating in an atmosphere with an exponential density profile given by
\begin{equation}
\rho=\rho_0 \exp[-z/h],
\end{equation}
where $\rho_0$ is the density on the ground, $z$ the height in the atmosphere measured from the ground and $h$ the atmospheric scale height. The atmosphere is assumed to be planar, which is valid for the terrestrial planets since their atmospheric scale heights are small compared to their radii. We further assume that radiative losses can be neglected such that the flow is adiabatic.
The adiabatic hydrodynamic equations are given by
\begin{equation}\label{e3}
\frac{1}{\rho}\frac{D\rho}{Dt}+\frac{\partial u}{\partial z}=0
\end{equation}
\begin{equation}
\frac{Du}{Dt}+\frac{1}{\rho}\frac{\partial p}{\partial z}=0
\end{equation}
\begin{equation}\label{e32}
\frac{1}{p}\frac{Dp}{Dt} - \frac{\gamma}{\rho}\frac{D\rho}{Dt}=0,
\end{equation}
where $\gamma$ is the adiabatic index and $D/Dt$ the ordinary Stokes time derivative.

Thanks to the self-similar behavior of the flow, the solutions to hydrodynamic equations above can be separated into their time-dependent and spatial parts and can be written as
\begin{equation}\label{e4}
\rho(z,t)=\rho_0 \exp[-Z(t)/h]G(\zeta), \quad  \ u(z,t)=\dot Z U(\zeta), \quad  \ p(z,t)=\rho_0 \exp[-Z(t)/h] \dot Z^2 P(\zeta)
\end{equation}
where $Z(t)$ is the position of the shock front and $\zeta=(z-Z(t))/h$. The similarity variables for the density, velocity and pressure are given by $G(\zeta)$, $U(\zeta)$ and $P(\zeta)$, respectively. Using the expressions in Equation (\ref{e4}) and substituting them into the hydrodynamic Equations (\ref{e3})-(\ref{e32}) yields for the spatial parts
\begin{equation}\label{e1011}
\frac{1}{G}\frac{dG}{d\zeta}(U-1)+\frac{dU}{d\zeta} =1
\end{equation}
\begin{equation}
(U-1)\frac{dU}{d\zeta}+\frac{1}{G}\frac{dP}{d\zeta}=-\frac{U}{\alpha}
\end{equation}
\begin{equation}\label{e1021}
(U-1)\left(\frac{1}{P}\frac{dP}{d\zeta}-\frac{\gamma}{G}\frac{dG}{d\zeta}\right)=-\frac{2}{\alpha}-\gamma+1,
\end{equation}
and a time dependent part given by
\begin{equation}\label{e103}
\frac{\dot Z^2}{\ddot Z}=\alpha h.
\end{equation}
Using the strong shock conditions we have
\begin{equation}\label{ein}
G(0)=\frac{\gamma+1}{\gamma-1}, \quad \ U(0)=\frac{2}{\gamma+1}, \quad \ P(0)=\frac{2}{\gamma+1}.
\end{equation}

Having separated the hydrodynamic equations into their time-dependent and spatial parts and we now obtain their self-similar solution. The solution to Equation (\ref{e103}) yields the position of the shock front as a function of time and is given by
\begin{equation}\label{e11}
Z(t)=-\alpha h \ln \left[1-\left(t/t_0\right)\right]
\end{equation}
where $t_0=2 \alpha h/[v_g (\gamma+1)]$ and $v_g$ is the ground velocity at the interface between the ground the atmosphere. Equation (\ref{e11}) shows that the shock accelerates fast enough such that it arrives at infinity in time $t_0$. The ground in contrast only transverses a distance $2\alpha h/(\gamma+1)$, which is a few scale heights, in the same time.

\begin{figure}[htp]
\centering
   \includegraphics[width=0.5\textwidth]{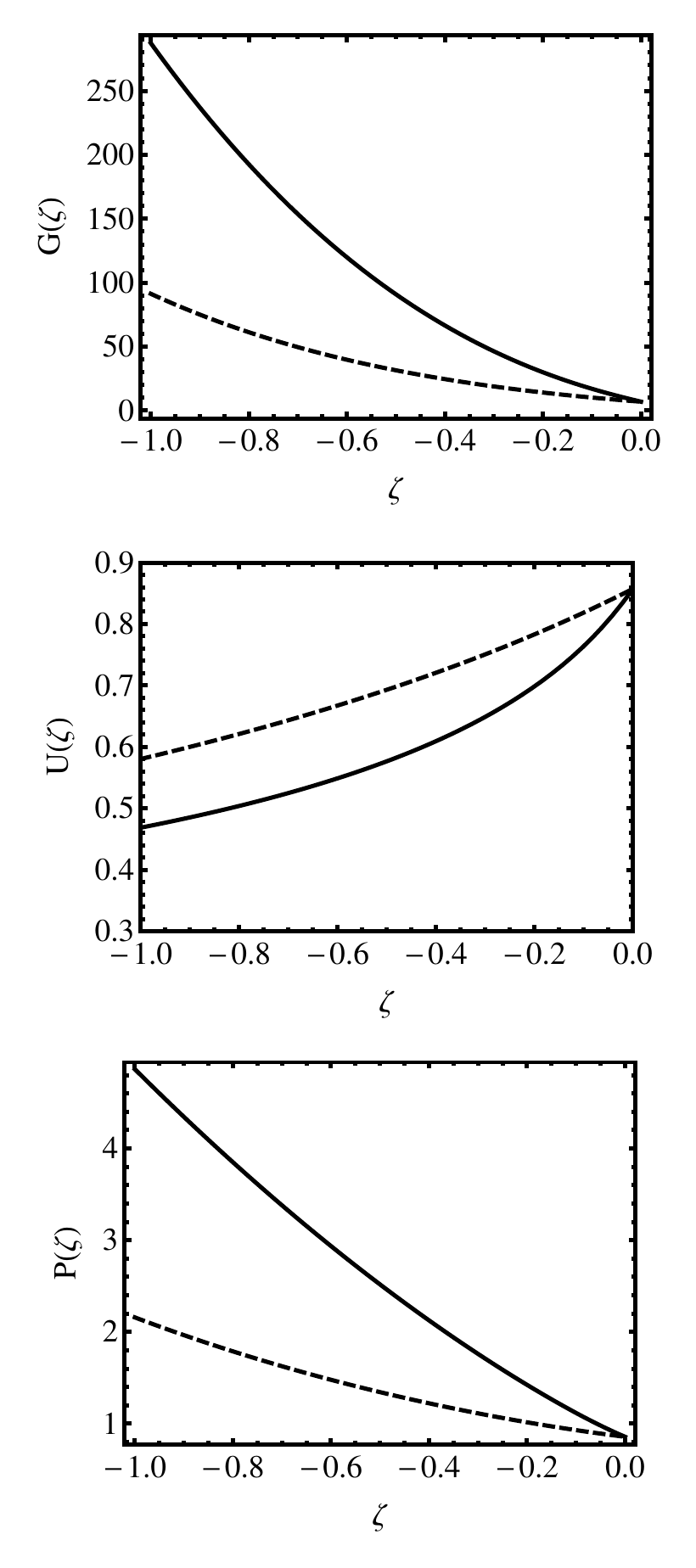}
\caption{Solutions for $G(\zeta)$, $U(\zeta)$ and $P(\zeta)$ for an adiabatic index $\gamma=4/3$ for an adiabatic atmospheric density profile, $\rho=\rho_0 (1-z/z_0)^{n}$ with $n=1.5$, (solid line) and an isothermal atmospheric density profile, $\rho=\rho_0\exp[-z/h]$, (dashed line).}
\label{fig2}
\end{figure}

Although various solutions to Equations (\ref{e1011})-(\ref{e1021}) exist for different values of $\alpha$, the physically relevant solution corresponds to a unique value of $\alpha$  which allows passage through a critical point, $\zeta_c$.  This critical point corresponds to  the sonic point in the time-dependent flow. Self-smilar solutions that include passage trough the sonic point are generally referred to as type II self-similar solutions. For example, we find, consistent with previous works \citep{GH66,C90}, that for $\gamma=4/3$, $\alpha=5.669$ and the critical point is located at $\zeta_c=-0.356$ and similarly, for $\gamma=5/3$, $\alpha=4.892$ and $\zeta_c=-0.447$. Since the self-similar solutions have to pass through the sonic point, only the region between the shock front and the sonic point are in communication and the part of the flow beyond the sonic point is cut off. The beauty of this is that the solution of the hydrodynamic equations becomes independent of the detailed nature of the initial shock conditions, such that the velocity of the ground motion that launches the shock only enters in the form of multiplicative constants in the asymptotic self-similar solution. Figure \ref{fig2} displays the solutions for $G(\zeta)$, $U(\zeta)$ and $P(\zeta)$ for an adiabatic index $\gamma=4/3$ for an isothermal atmospheric density profile and adiabatic atmospheric density profile (see section 2.2).

\begin{figure}[htp]
\centering
   \includegraphics[width=1.0\textwidth]{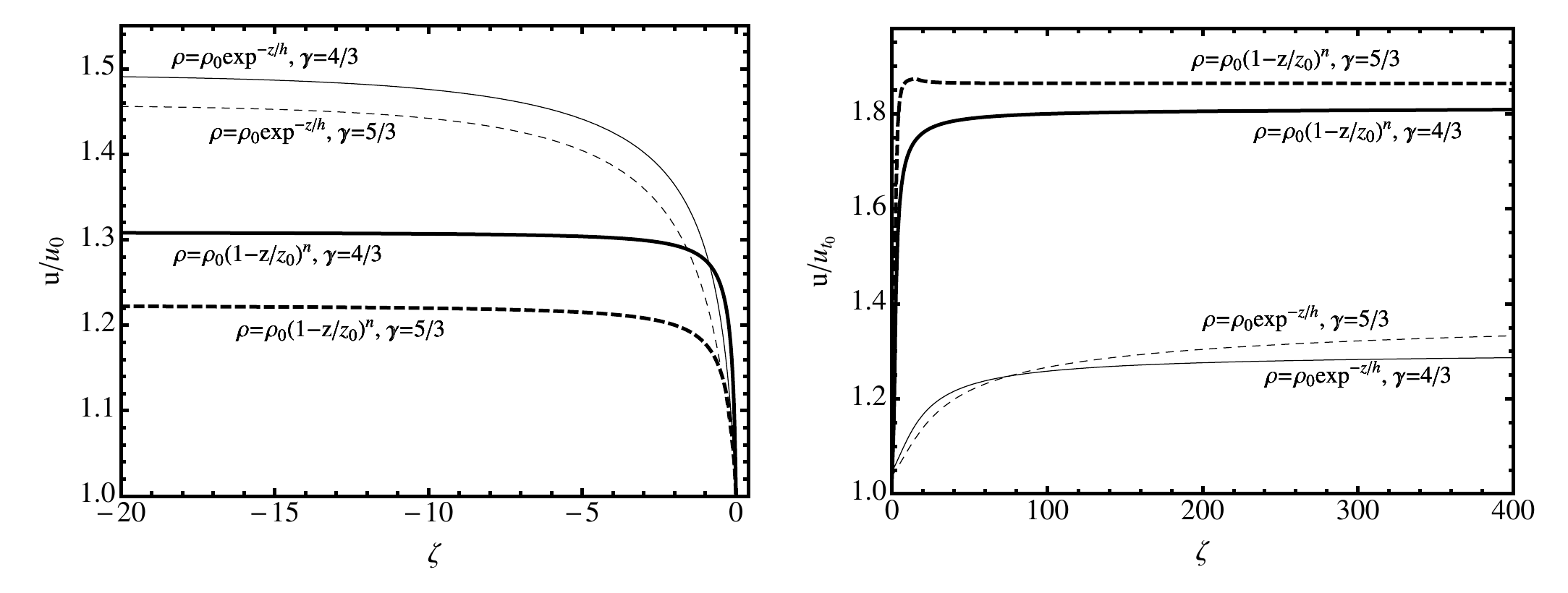}
\caption{Left: $u/u_0$ as a function of distance form the shock front, $\zeta$, for an isothermal (thin lines) and adiabatic density profile (thick lines). The solid and dashed lines correspond to $\gamma=4/3$ and $\gamma=5/3$, respectively. The value of acceleration factor until the shock reached $\infty$, $u_{t_0}/u_{0}$, can be read of the left side of the figure corresponding to large distances from the shock front. Right: $u/u_{t_0}$ as a function of $\zeta$. The value of acceleration factor from the time the shock reached $\infty$ until a long time after that, $u_{\infty}/u_{t_0}$, can be read of the right side on the figure corresponding to late times long after the shock front reached $\infty$. The total acceleration factor, $\beta$, is the product of the $u_{t_0}/u_{0}$ shown in the left Figure and  $u_{\infty}/u_{t_0}$ shown in the right Figure.}
\label{fig3}
\end{figure}

The atmospheric mass loss fraction for an exponential atmosphere is
\begin{equation}\label{e1}
\chi_{loss}=\exp[-z_{esc}/h]
\end{equation}
where $z_{esc}$ is the initial height in the atmosphere of the fluid element that has a velocity equal to the escape velocity at a time long after the shock has passed, such that the atmosphere at $z \geq z_{esc}$ will be lost. From Equation (\ref{e11}) we have that the shock velocity grows exponentially with height in the atmosphere just as the density deceases exponentially. The shock velocity is given by
\begin{equation}
\dot Z=\frac{\gamma+1}{2}v_g \exp[z/\alpha h].
\end{equation}
$z_{esc}$ can therefore be written as $v_{esc}=v_{g} \beta \exp[z_{esc}/\alpha h]$ where $v_{esc}$ is the escape velocity of the impacted body and $\beta$ is a numerical constant that relates the velocity of a given fluid element at a time long after the shock has passed, $u_{\infty}$, to the velocity of the same fluid element at the shock, $u_0$.
The final atmospheric mass loss fraction is therefore
\begin{equation}\label{e11X}
\chi_{loss}=\left( \frac{\beta v_g}{v_{esc}} \right) ^\alpha,
\end{equation}
where the only quantity left to calculate numerically is the acceleration factor $\beta$ given by
\begin{equation}\label{e51}
\beta = \frac{u_{\infty}}{u_0}.
\end{equation}
It is convenient to write $\beta$ as the product of the acceleration factor until the shock has reached infinity $(u_{t_0}/u_{0})$, which happens at time $t_0$, and the acceleration factor from the time that the shock reached $\infty$ to a long time after that $(u_{\infty}/u_{t_0})$, such that $\beta = (u_{\infty}/u_{t_0})(u_{t_0}/u_{0})$. The latter is important because a given fluid element continues to accelerate after $t_0$. The two parts of the acceleration factor can be written as
\begin{equation}\label{eb1}
\frac{u_{t_0}}{u_0}=\frac{U(\zeta \to -\infty) \dot Z(\zeta \to -\infty)}{U(\zeta =0) \dot Z(\zeta=0)}, \quad \frac{u_{\infty}}{u_{t_0}}=\frac{U(\zeta \to +\infty) \dot Z(\zeta \to +\infty)}{U(\zeta \to -\infty) \dot Z(\zeta \to -\infty)}.
\end{equation}
Equations (\ref{eb1}) required that we take the limit for $\zeta$ and $t$ together. This is accomplished by rewriting $\dot Z$ as $d \ln \dot Z/d \zeta= (\alpha(U(\zeta)-1))^{-1}$ and solving it together with Equations (\ref{e1011}) - (\ref{e1021}). Figure \ref{fig3} displays the two components of the acceleration factor and we find that $\beta=2.07$ for $\gamma=4/3$ and $1.90$ for $\gamma=5/3$.

\begin{figure}[htp]
\centering
   \includegraphics[width=0.7\textwidth]{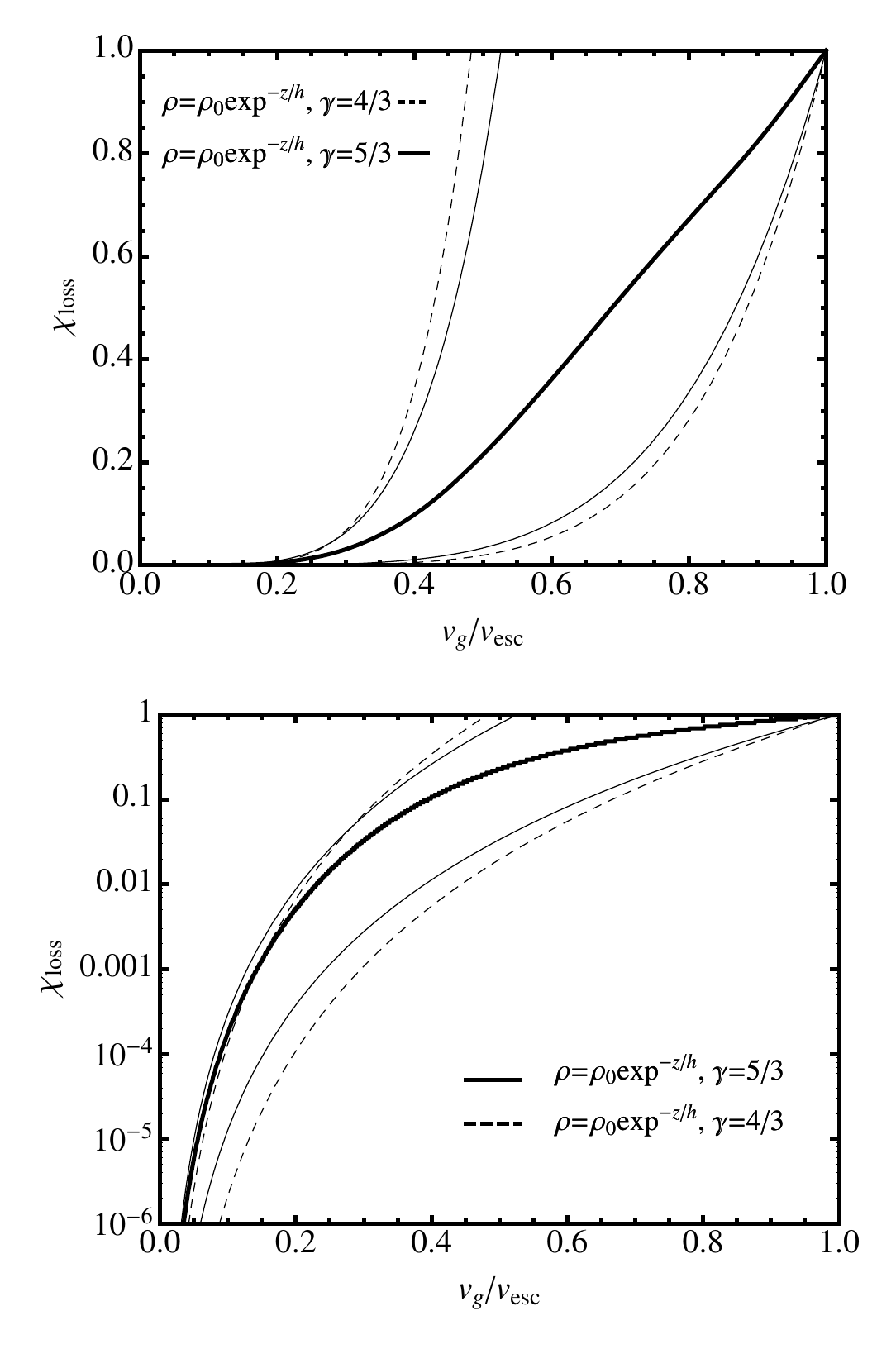}
\caption{Mass Loss Fraction, $\chi_{loss}$, as a function of $v_{g}/v_{esc}$ for an isothermal atmosphere. The thin solid and thin dashed lines correspond to $\gamma=5/3$ and $\gamma=4/3$, respectively. Self-similar solutions with $\beta=1$ correspond to the lower two curves and with $\beta=2.07$ ($\gamma=4/3$) and $\beta=1.90$ ($\gamma=5/3$) to the upper curve two curves, respectively. The thick black line represents the atmospheric mass loss fraction obtained from full numerical integrations for $\gamma=5/3$.}
\label{fig4}
\end{figure}

Because the shock is not immediately self-similar from the very moment that it is launched into the atmosphere, the actual acceleration factor, $\beta$, is less than the value of $\beta$ obtained from the self-similar solutions. Furthermore, the atmosphere close to the ground is not accelerated as much as fluid elements with initial positions significantly above the ground. Therefore, in order to obtain the actual value of $\beta$ and an accurate atmospheric mass loss for the part of the atmosphere that resides close to the ground, we performed full numerical integrations of the hydrodynamic equations. The simulations were performed using the one dimensional version of RICH (Yalinewich et al., in preparation), a Godunov type hydro-code on a moving Lagrangian mesh. We used a grid with a total of 1000 elements and as boundary conditions we used a piston moving at a constant velocity on one side and assumed a vacuum on the other. Due to numerical reasons, we couldn't set the initial upstream pressure to zero, so we used a small value of $10^{-9}$. We verified that the results converged by running the same simulation with half as many grid points. Figure \ref{fig4} shows the atmospheric mass loss fraction, $\chi_{loss}$, as a function of $v_{g}/v_{esc}$ from our self-similar solutions (thin lines) with $\beta=1$ (lower curves) and $\beta=2.07$ ($\gamma=4/3$, dashed upper curve) and $\beta=1.90$ ($\gamma=5/3$, solid upper curve). The numerical solution for $\gamma=5/3$ is represented by the thick line. As expected, the full numerical solution for $\gamma=5/3$ falls between the $\beta=1$ and $\beta=1.90$ lines and we find that the actual value of $\beta$ is 1.71.

Given the resulting distribution of ground velocities, $v_g$, from a giant impact (see section 2.3), Equation (\ref{e11X}) can be used to determine the global atmospheric mass loss fraction for an isothermal atmosphere.

\subsection{Self-Similar Solutions to the Hydrodynamic Equations for an Adiabatic Atmosphere}
The heat transport in many of the close-in exoplanet atmosphere may be dominated by convection rather than radiation, resulting in adiabatic atmospheres. Unlike an isothermal atmosphere, an adiabatic atmosphere has a density profile that reaches $\rho=0$ at a finite distance from the planet. 
Similar to the isothermal density profile considered above, we can repeat our calculation for an atmosphere with an adiabatic density profile given by
\begin{equation}
\rho=\rho_0 (1-z/z_0)^{n},
\end{equation}
where $z_0$ is the edge of the atmosphere where $\rho=0$ and $P=0$ and $n$ is the polytropic index. We again assume that the atmosphere is planar and that radiative losses can be neglected such that the flow is adiabatic. 
For the adiabatic density profile the solutions to hydrodynamic equations above can again be separated into their time-dependent and spatial parts and are given by
\begin{equation}\label{e44}
\rho(z,t)=\rho_0 (1-Z(t)/z_0)^{n}G(\zeta), \quad  \ u(z,t)=\dot Z U(\zeta), \quad  \ p(z,t)=\rho_0 (1-Z(t)/z_0)^{n} \dot Z^2 P(\zeta)
\end{equation}
where $Z(t)$ is the position of the shock front and $\zeta=(z-Z(t))/(z_0-Z(t))$. 

Using the expressions in Equation (\ref{e44}) and substituting them into the hydrodynamic Equations (\ref{e3})-(\ref{e32}) yields for the spatial parts
\begin{equation}\label{ea1}
\frac{1}{G}\frac{dG}{d\zeta}(U-1+\zeta)+\frac{dU}{d\zeta} =n
\end{equation}
\begin{equation}
(U-1+\zeta)\frac{dU}{d\zeta}+\frac{1}{G}\frac{dP}{d\zeta}=-\frac{U}{\alpha}
\end{equation}
\begin{equation}\label{ea2}
(U-1+\zeta)\left(\frac{1}{P}\frac{dP}{d\zeta}-\frac{\gamma}{G}\frac{dG}{d\zeta}\right)=-\frac{2}{\alpha}-(\gamma-1)n,
\end{equation}
and a time dependent part given by
\begin{equation}\label{e1033}
\frac{\dot Z^2}{\ddot Z (1-Z/z_0)}=\alpha z_0.
\end{equation}
Solving Equation (\ref{e1033}) using the same strong shock initial conditions given in Equation (\ref{ein}) yields for the position of the shock front as a function of time
\begin{equation}\label{e1034}
Z(t)=z_0\left[1-\left(1-\frac{t}{t_0}\right)^\frac{\alpha}{1+\alpha}\right]
\end{equation}
where $t_0=2 z_0 \alpha/(v_g (1+\alpha) (1+\gamma))$ is the time at which the shock reaches the edge of the atmosphere at $z=z_0$.

Just like for the exponential atmosphere, the physically relevant solution to Equations (\ref{ea1}) - (\ref{ea2}) for the adiabatic atmosphere density profile corresponds to a unique value of $\alpha$ which allows passage through the critical point. We find, for $\gamma= 4/3$, $\alpha=1.796$ and $\zeta_C=-0.083$ and for $\gamma=5/3$, $\alpha=3.029$ and $\zeta_C=-0.156$. Figure \ref{fig2} shows the solutions for $G(\zeta)$, $U(\zeta)$ and $P(\zeta)$ for $\gamma=4/3$ for an adiabatic atmospheric density profile (solid line) and an isothermal atmospheric density profile (dashed line).

The atmospheric mass loss fraction for an adiabatic atmosphere is
\begin{equation}
\chi_{loss}=\left( 1-\frac{z_{esc}}{z_0}\right)^{n+1}
\end{equation}
where $z_{esc}$ is the initial height in the atmosphere of the fluid element that has a velocity equal to the escape velocity at a time long after the schlock has passed. From Equation (\ref{e1034}) we have that the shock accelerates with height in the atmosphere and the shock velocity is given by
\begin{equation}
\dot Z=\frac{\gamma+1}{2}v_g \left(1-\frac{z}{z_0} \right)^{-1/\alpha}.
\end{equation}
$z_{esc}$ can therefore be written as  $v_{esc}=v_{g} \beta (1-z_{esc}/z_0)^{-1/\alpha}$. $\beta$ is again a numerical constant that relates the velocity of a given fluid element at a time long after the shock has passed, $u_{\infty}$, to the velocity of the same fluid element at the shock, $u_0$.
The final atmospheric mass loss fraction is therefore
\begin{equation}\label{e1901}
\chi_{loss}=\left( \frac{\beta v_g}{v_{esc}} \right) ^{\alpha (n+1)}.
\end{equation}
Calculating $\beta$ using an analogous procedure to one employed for the isothermal atmosphere in section 2.1 with the main difference that $\dot Z$ is now given by $d \ln \dot Z/d \zeta= (\alpha(U(\zeta)-1+\zeta))^{-1}$, we find $\beta=2.38$ and $\beta=2.27$ for for $\gamma=4/3$ and $\gamma=5/3$, respectively. Figure \ref{fig3} shows the two components of the acceleration as a function of the distance from the shock front, $\zeta$, for an exponential and adiabatic atmospheric density profile.

\begin{figure}[htp]
\centering
   \includegraphics[width=0.7\textwidth]{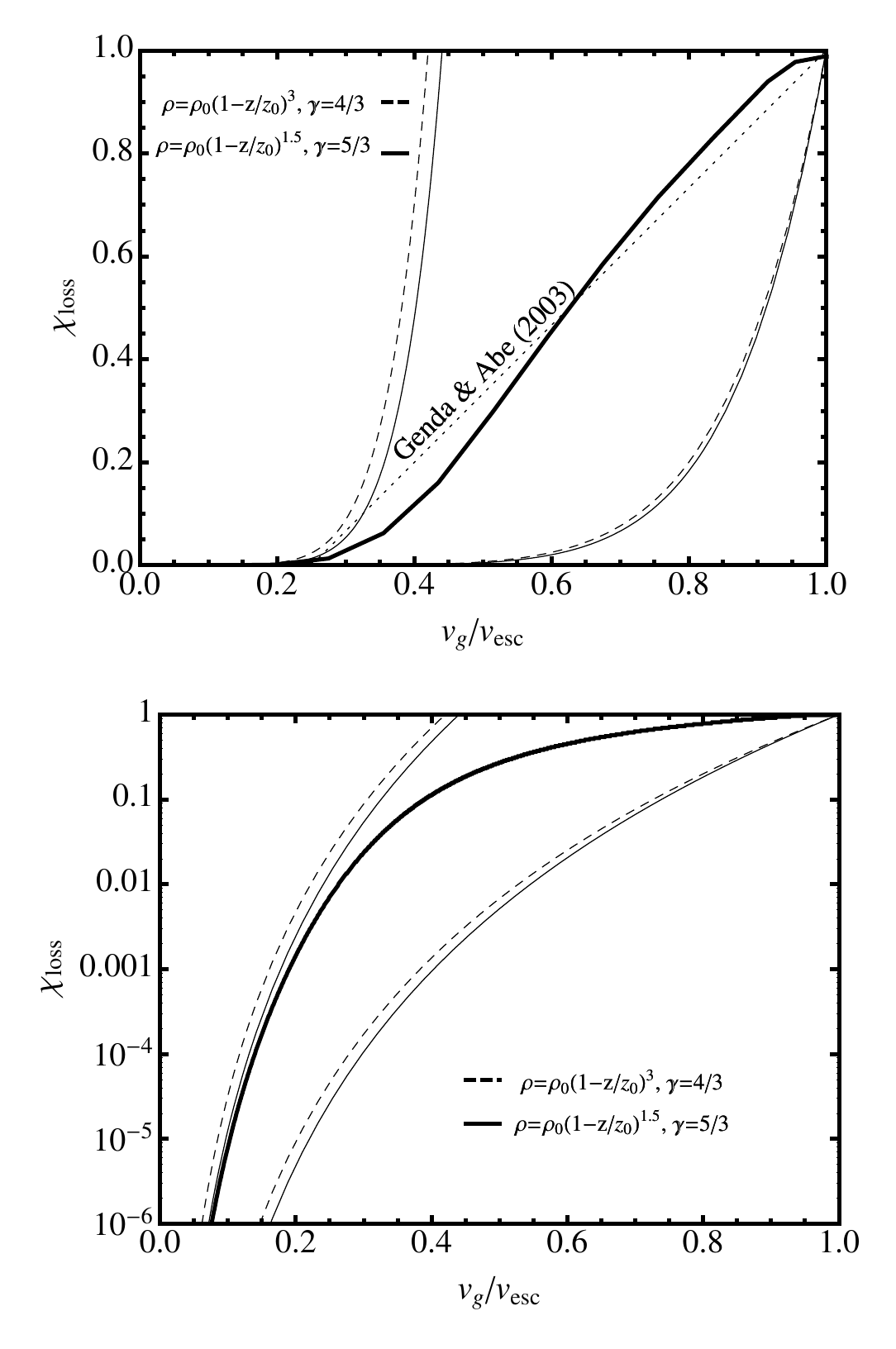}
\caption{Same is in Figure \ref{fig4} but for an adiabatic atmosphere. The dotted line represent the atmospheric mass loss results from \citet{GA03}.}
\label{fig5}
\end{figure}

Therefore, the exponent of $\beta v_g/v_{esc}$ for an adiabatic atmosphere is, for example, 7.2 for $n=3$ and $\gamma=4/3$ and 7.6 for $n=1.5$ and $\gamma=5/3$ compared to $5.7$ ($\gamma=4/3$) and $4.9$ ($\gamma=5/3$) for an isothermal atmosphere, respectively. Figure \ref{fig5} shows the fractional atmospheric mass loss as a function of the ground velocity, $v_g$, as obtained from our analytic self-similar solutions and full numerical integrations.

Equation (\ref{e1901}) gives the local atmospheric mass loss fraction for an adiabatic atmosphere as a function of ground velocity. To obtain the global atmospheric mass loss due to a giant impact, one needs to obtain the resulting distribution of ground velocities, $v_g$, across the planet from a giant impact and use these to calculate the local atmospheric mass loss and sum the results over the whole planet. In the following subsection (section 2.3) we use a simple impact model to obtain the global atmospheric mass loss as a function of the impactor mass and velocity.

\subsection{Global Atmospheric Mass Loss}
\subsubsection{Relating the Global Ground Motion to Impactor Mass and Velocity}
\begin{figure}[htp]
\centering
   \includegraphics[width=0.7\textwidth]{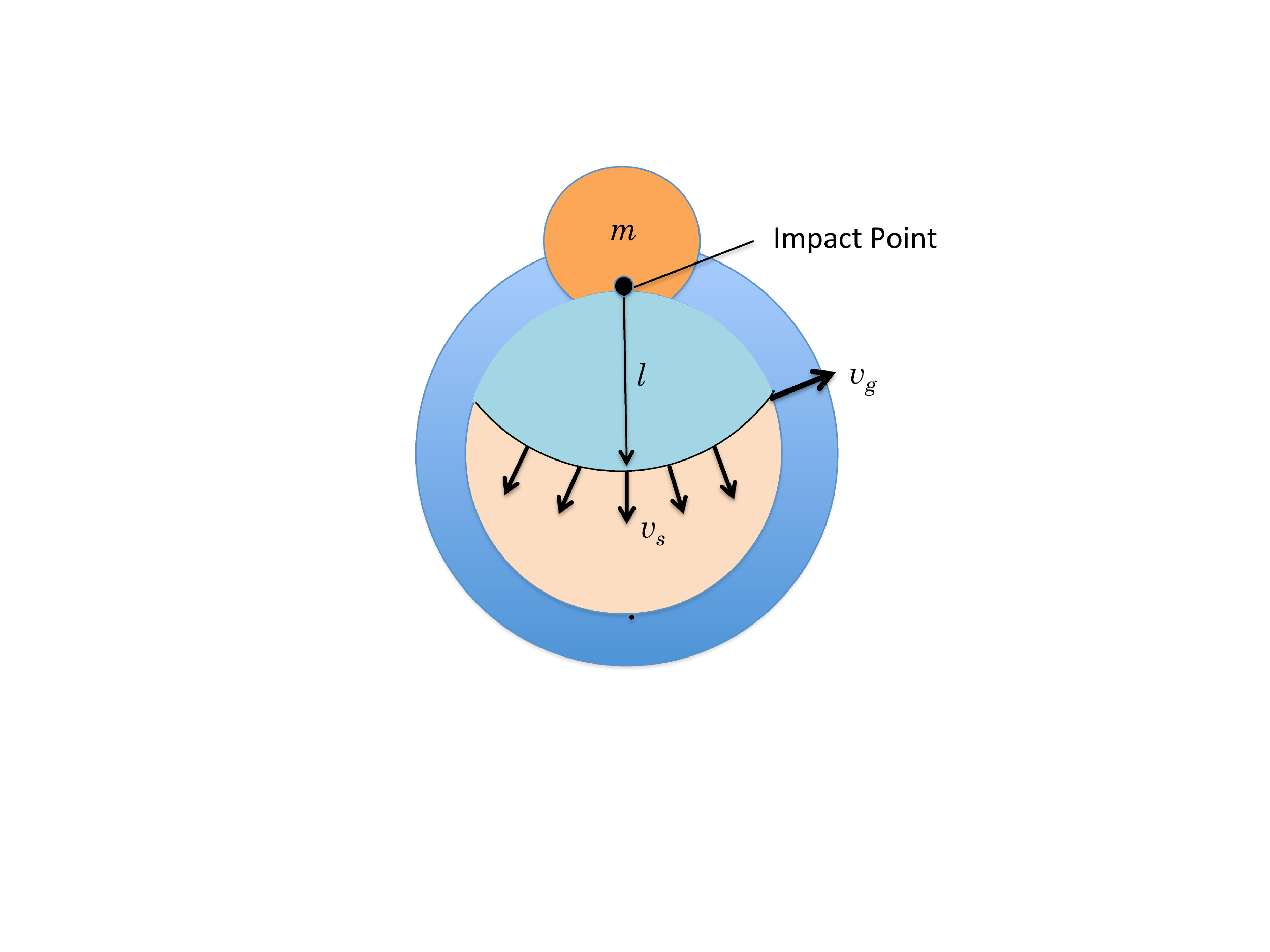}
\caption{Illustration of the impact geometry. An impactor of mass, $m$, and impact velocity, $v_{imp}$, impacts a target with mass, $M$, and radius, $R$. Assuming momentum conservation, we calculate the shocked fluid velocity, $v_s$, and the component of the ground velocity normal to the surface, $v_{g}$, as a function of the distance from the impact point.}
\label{fig16}
\end{figure}

\begin{figure}[htp]
\centering
   \includegraphics[width=0.7\textwidth]{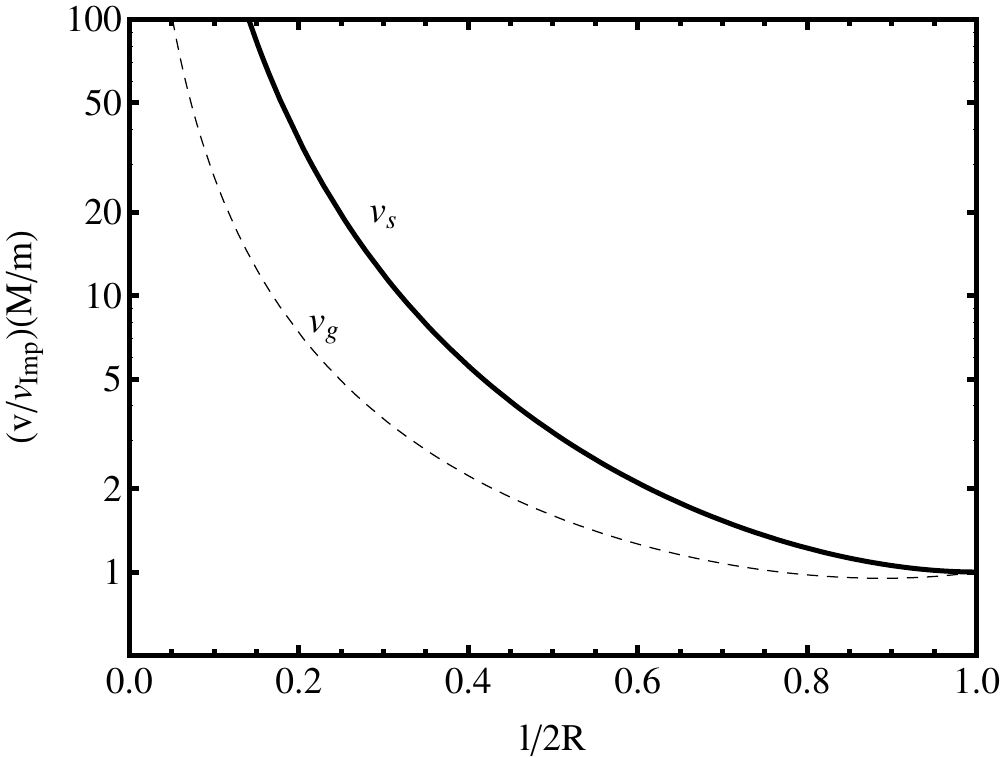}
\caption{Shocked fluid velocity $v_s$ and the ground velocity $v_g$ as a function of distance travelled by the shock, $l$, from the impact point to the other side of the planet, $l=2R$.}
\label{fig6}
\end{figure}

To obtain the total atmospheric mass lost in a given impact we need to relate the impactor mass, $m$, and impact velocity, $v_{Imp}$, to the resulting ground motion at the various locations of the protoplanet and use these together with Equations (\ref{e11X}) and (\ref{e1901}) to obtain the local atmospheric mass loss and sum the results over the surface of the planet. When an impactor hits a protoplanet, it initially transfers most of its energy to a volume comparable to its own size at the impact site. A significant fraction of this energy will escape from the site via a small amount of impact ejecta, but some of the energy will propagate through the protoplanet as a shock. Using a very simple impact model, we approximate the impacts as point like explosions on a sphere. This is similar to the treatment of point like explosions on a planar surface between a vacuum and a half-infinite space filled with matter. Such an explosion results again in a self-similar solution of the second type \citep{ZR67}. As the shock propagates it must lose energy because some of the shocked material flows into vacuum, but its momentum is increased by the nonzero pressure in the protoplanet. As a result, the shock's velocity should fall off faster than dictated by energy conservation but slower than required by momentum conservation. Numerical simulations of catastrophic impacts find scaling laws that are close to the ones derived by assuming momentum conservation \citep{LA96,BA99}. For example, \citet{LA96} find that the catastrophic destruction threshold, defined as the impact energy per unit target mass required to eject 50\% of the target, scales as $R^{1.1}$, which is close to the linear scaling with $R$ predicted from momentum conservation for fixed impactor velocity. We therefore assume momentum conservation of the shock, $m v_{imp}=M v_s$, as it propagates through the target and use it to calculate the resulting ground velocity across the protoplanet (see Figure \ref{fig16}). This treatment is similar to the `snowplow' phase of an expanding supernova remnant during which the matter of the ambient intersteller medium is swept up by the expanding shock and momentum is conserved. The volume of the protoplanet that a spherical shock, originating from an impact point on the protoplanet's surface, transversed as a function of distance from the impact point, $l$, is given by $V=\pi l^3(4-3(l/2R))/6$ and shown as light blue region in Figure \ref{fig16}. This volume is equivalent to the volume of two intersecting spheres with radii $R$ and $l$ where the center of sphere corresponding to the shock co-insides with the surface of the protoplanet of radius, $R$. Assuming a constant density of the target and momentum conservation the velocity of the shocked fluid traveling through the protoplanet is given by
\begin{equation}\label{eaxr1}
v_s=v_{Imp} \left(\frac{m}{M}\right)\frac{1}{(l/2R)^3(4-3(l/2R))},
\end{equation}
where $l$ is the distance of the shock travelled from the impact point, such that $l=2R$ when the shock reaches the antipode (see Figure \ref{fig16}). The ground velocity with which the shock is launched into the atmosphere is due to the component of the shocked fluid velocity that is perpendicular to the planet's surface, such that $v_g=v_s l/(2R)$, which yields
\begin{equation}\label{eax1}
v_g=v_{Imp} \left(\frac{m}{M}\right)\frac{1}{(l/2R)^2(4-3(l/2R))}.
\end{equation}
Figure \ref{fig6} shows the shocked fluid velocity, $v_s$, and the ground velocity, $v_g$, as a function of distance travelled by the shock through the planet. $v_g$ has a minimum at $l/2R=8/9$. Our simple impact model assumes that the target has a constant density and neglects any impact angle dependence. The latter is a reasonable assumption as long as the impactor mass is significantly less than the target mass. The former is a reasonable first order approximation given our general ignorance concerning the interior structure of planetary embryos during their formation.

Equations (\ref{eaxr1}) and (\ref{eax1}) assume momentum conservation within the target. For comparison, if we instead assume momentum conservation in a uniform density half-infinite sphere and compare it to Equations (\ref{eaxr1}) and (\ref{eax1}), we find average shocked fluid velocities and ground velocities that are about a factor of 2 smaller. This implies that we may somewhat overestimate the global atmospheric loss due to giant impacts.

\subsubsection{Global Atmospheric Mass Loss Results}
To ensure the entire atmosphere is lost we required that $v_g(l/2R=8/9) \geq v_{esc}$ (see Equations (\ref{e11X}), (\ref{e1901}) and (\ref{eax1})), where we set $\beta=1$ to account for the fact that if we want to eject all of the atmosphere, we do have to lose also the part of the atmosphere immediately above the ground for which $\beta=1$.
Substituting for $v_g$ and rearranging yields that all of the atmosphere is lost provided that 
\begin{equation}
\left(\frac{v_{Imp}}{v_{esc}}\right)\left(\frac{m}{M}\right)\frac{243}{256} \geq 1.
\end{equation}

\begin{figure}[htp]
\centering
    \includegraphics[width=0.7\textwidth]{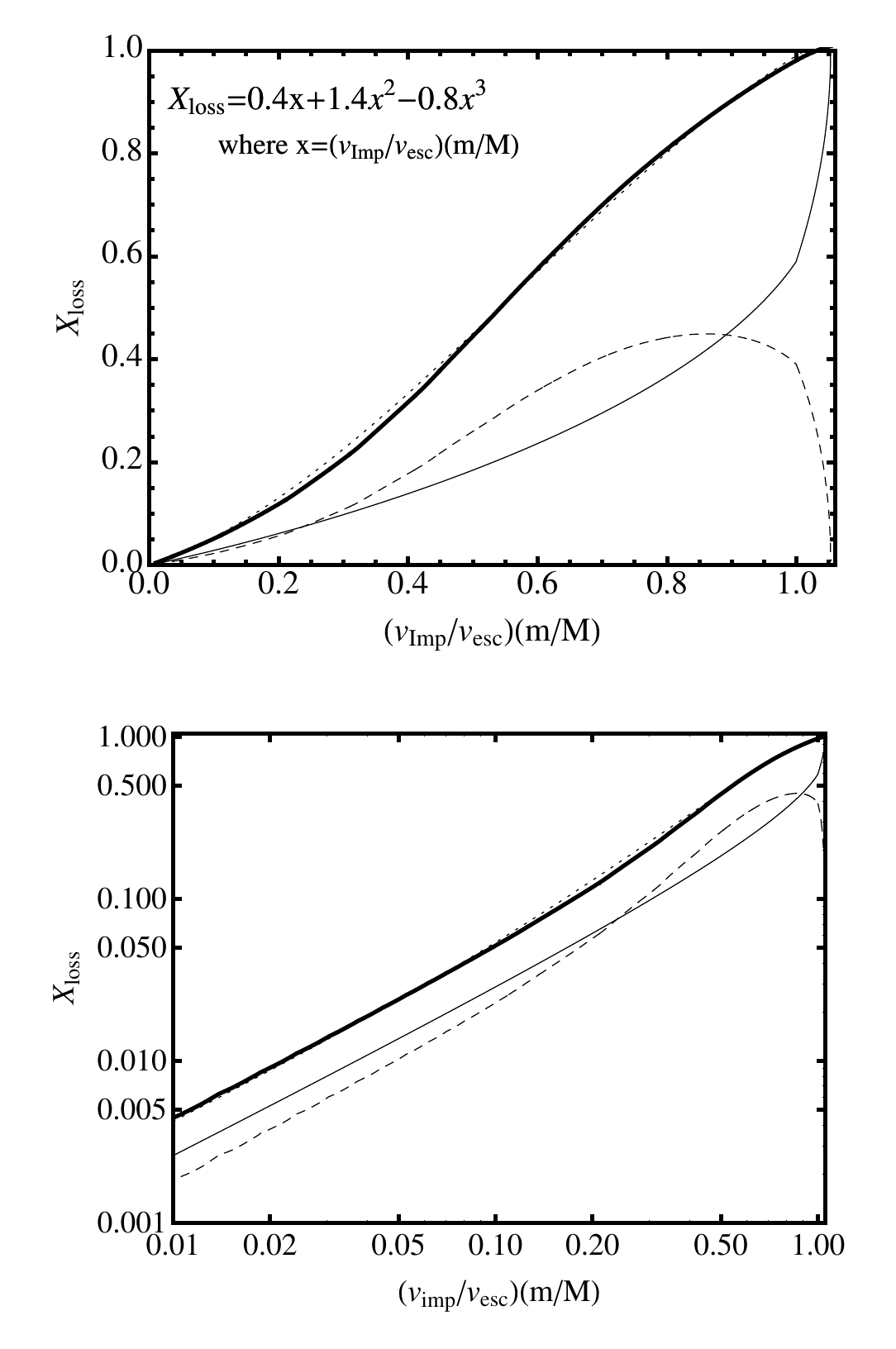}
\caption{Global mass loss fraction (thick solid line), calculated by taking into account the different ground velocities across the planet's surface. The total atmospheric loss consists of two components: The first is from the area of of the planet's surface where the ground motion is large enough such that locally all the atmosphere is lost (dashed line) and the second component corresponds to the regions of the planet's surface where the local ground velocity is small such that only part of the atmosphere is lost (thin solid line). A good fit over the whole range of $(v_{Imp}/v_{esc})(m/M)$ is given by $X_{loss}=0.4(v_{Imp}/v_{esc})(m/M)+1.4(v_{Imp}/v_{esc})(m/M)^2-0.8(v_{Imp}/v_{esc})(m/M)^3$ (dotted line).}
\label{fig7}
\end{figure}

Only part of the global atmosphere is lost for
\begin{equation}
\left(\frac{v_{Imp}}{v_{esc}}\right) \left(\frac{m}{M}\right) \frac{243}{256} < 1.
\end{equation}
The atmospheric mass loss as a function of $(v_{Imp}/v_{esc})(m/M)$ is shown in Figure \ref{fig7}. When only a fraction of the atmosphere is lost, it is interesting to note that the total atmospheric loss consists of two components: The first is from the area of the planet's surface where the ground motion is large enough such that locally all the atmosphere is lost (dashed line in Figure \ref{fig7}), the second component corresponds to the region of the planet where the local ground velocity is small enough such that only part of the atmosphere is lost (thin solid line in Figure \ref{fig7}). In the latter case, the local fractional mass loss is given by Equation (\ref{e11X}) for an isothermal and Equation (\ref{e1901}) for an adiabatic atmosphere, respectively. 

In the limit that $(v_{Imp}/v_{esc})(m/M) \ll 1$, Equation (\ref{eax1}) simplifies to $v_{esc}=v_{Imp}(m/4M)(2R/l)^2$ such that in the limit of small total atmospheric mass loss we have
\begin{equation}\label{gi0}
X_{loss} = \left( \frac{l}{2R}\right)^2 \simeq \left( \frac{m}{4M}\right) \left(\frac{v_{Imp}}{v_{esc}}\right). 
\end{equation}
In addition to the regions undergoing total atmospheric loss, we also have a contribution form parts of the planet undergoing partial loss, yielding a total atmospheric mass loss fraction $X_{loss} = 0.4 (m/M)(v_{Imp}/v_{esc})$. We note here that this formalism is less accurate for small impactor masses with $v_{Imp} \sim v_{esc}$, since it does not include any atmosphere ejected directly at the impact site (see Section 3).

More generally, we find that the global mass loss fraction for an isothermal atmosphere is, independent of the exact value of the adiabatic index, well approximated by 
\begin{equation}\label{aml1}
X_{loss}=0.4 \left(\frac{v_{Imp} m}{v_{esc} M}\right)+1.4\left(\frac{v_{Imp} m}{v_{esc} M}\right)^2-0.8\left(\frac{v_{Imp} m}{v_{esc} M}\right)^3
\end{equation}
and is plotted as dotted line, which is barely distinguishable from the thick solid line, in Figure \ref{fig7}.

Similarly, for an adiabatic atmosphere we find
\begin{equation}\label{aml2}
X_{loss}=0.4 \left(\frac{v_{Imp} m}{v_{esc} M}\right)+1.8 \left(\frac{v_{Imp} m}{v_{esc} M}\right)^2-1.2 \left(\frac{v_{Imp} m}{v_{esc} M}\right)^3.
\end{equation}

\begin{figure}[htp]
\centering
    \includegraphics[width=0.8\textwidth]{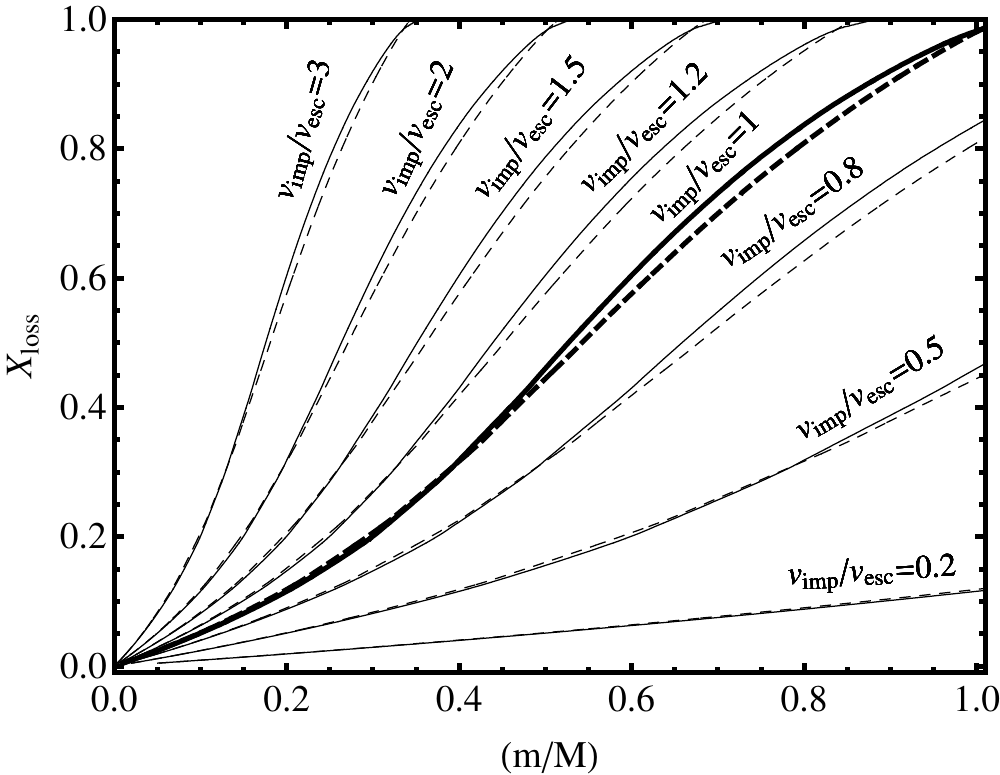}
\caption{Global mass loss fraction for an isothermal atmosphere (solid lines) and an adiabatic atmosphere (dashed lines) as a function of impactor mass to target mass ratio, $m/M$, calculated by taking into account the different ground velocities across the planet's surface.}
\label{fig8}
\end{figure}

Figure \ref{fig8} shows the total atmospheric mass loss fraction for an isothermal (solid lines) and adiabatic atmosphere (dotted line) as a function of impactor to target mass ratio for various impact velocities. 
For a Mars-sized impactor hitting an $0.9~M_{\earth}$ protoplanet with $v_{Imp} \sim v_{esc}$, we find $X_{loss}=6$\%. This is about a factor of 2 lower than estimates by \citet{GA03} who assumed an average ground velocity of $4-5$~km/s across the whole protoplanet and used this velocity together with their local atmospheric mass loss results (similar to the ones shown in Figure \ref{fig5}) to estimate a global atmospheric mass loss of 10\%. We show here, however, that the global atmospheric mass loss consists of two components, where the first component is from parts of the planet where the ground motion is large enough such that locally all the atmosphere is lost (dashed line in Figure \ref{fig7}) and the second component corresponds to the region of the planet where the local ground velocity is small enough such that only part of the atmosphere is lost (thin solid line in Figure \ref{fig7}). This makes the {\it average} ground velocity inadequate for determining the global atmospheric mass loss.

In the atmospheric mass loss calculations presented in this section, we assume that ratio of specific heats, $\gamma$,  is constant throughout the flow. However, the temperatures reached during the shock propagation are high enough to lead to ionization of the atmosphere, which in turn will decrease the value of $\gamma$ and consequently result in reduced atmospheric mass loss. The atmospheric mass loss due to giant impacts calculated in this section is therefore an overestimate.

\section{Atmospheric Mass Loss Due to Planetesimal Accretion and the Late Veneer}\label{s2}
Although smaller impactors cannot individually eject a large fraction of the planetary atmosphere, they collectively can play an important role in atmospheric erosion and, as we show in section 4, may easily dominate atmospheric mass loss during planet formation.

\subsection{Planetesimal Impacts}
\begin{figure}[htp]
\centering
    \includegraphics[width=1.0\textwidth]{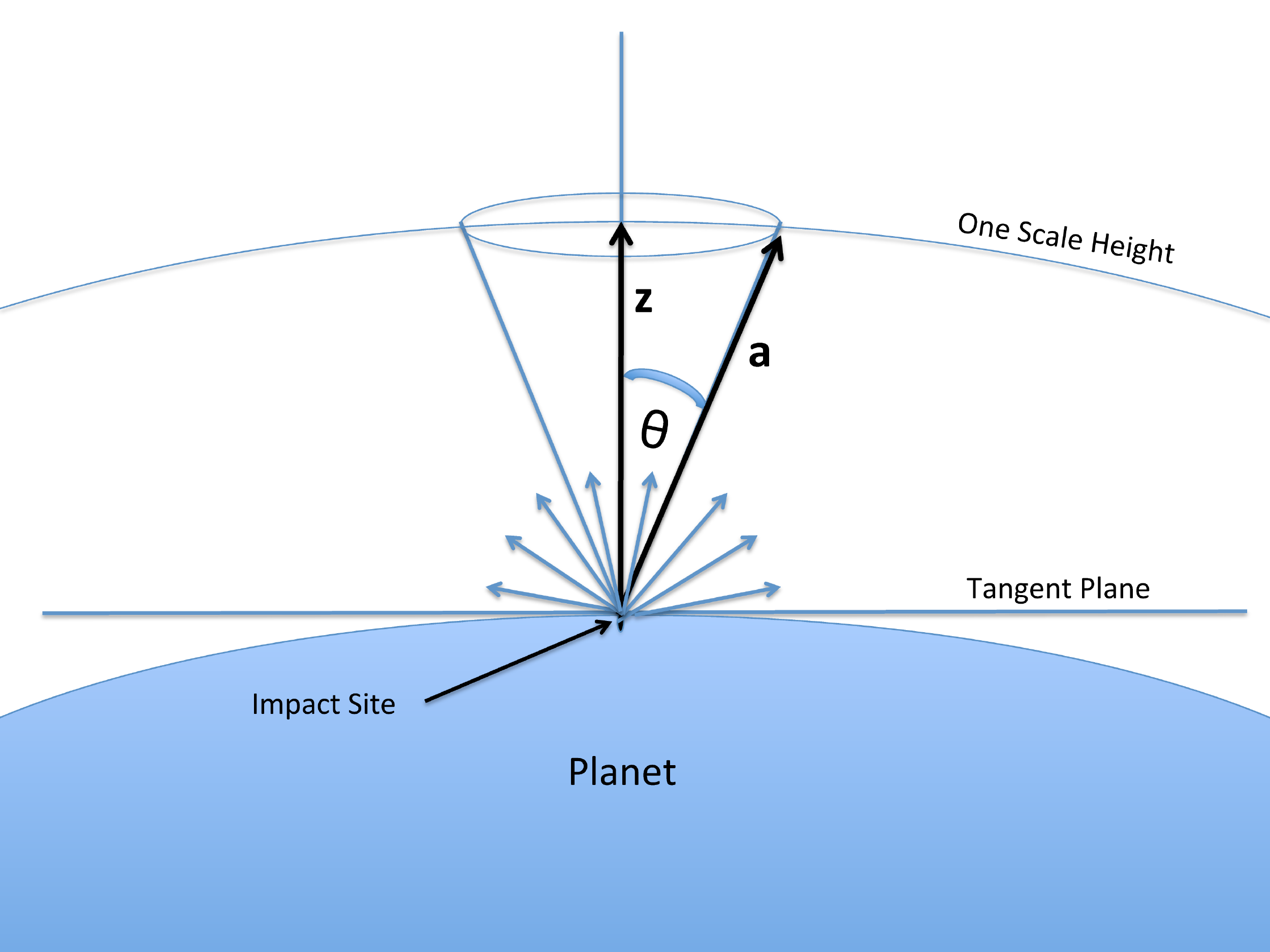}
\caption{Illustration of the impact geometry. Planetesimal impacts can only eject atmosphere locally. Treating their impact as a point-like explosion leading to an isotropic shock at the impact site, the maximum atmospheric mass that they can eject in a single impact is given by all the mass above the tangent plane, which is  $h/2R$ of the total atmosphere. However, since smaller impactors are more numerous than larger ones  required for giant impacts, smaller impactors may actually dominate the atmospheric mass loss during planet formation.}
\label{fig14}
\end{figure}

Unlike giant impacts which can create a strong shock propagating through the planetary interior that in turn can launch a strong shock into the planetary atmosphere, smaller planetesimal collisions can only eject the atmosphere locally. When a high-velocity impactor hits the surface of the protoplanet, its velocity is sharply decelerated and its kinetic energy is rapidly converted into heat and pressure resulting in something analogous to an explosion \citep{ZR67}. Similar to \citet{VM90}, we model the impact as a point explosion on the surface, where a mass equal to the mass of the impactor, $m_{Imp}$, propagates isotropically into a half-sphere with velocity of order, $v_{esc}$.
Atmosphere is ejected only where its mass per unit solid angle, as measured from the impact point, is less than that of the ejecta, $m_{Imp}/2\pi$.
We can then relate the impactor mass, $m_{Imp}$, to the ejected atmospheric mass $\mathcal{M}_{eject}$ (see following Equations (\ref{sp1}), (\ref{e101}) and (\ref{ex2})). These two masses are not equal because the planetesimal impact launches a point-like isotropic explosion into a half-sphere on the planetary surface, but the atmospheric mass above the tangent plane is not isotropically distributed around the impact site (see Figure \ref{fig14}), but is more concentrated towards the horizon. Specifically, the atmospheric mass close to the tangent plane of the impact site is hardest to eject due to its larger column density.

In order to distinguish between the impactor mass and the mass ejected from the atmosphere we use $\mathcal{M}$ for the mass in the atmosphere that is ejected and, as in section 2, $m$ and $r$ to describe the mass and radius of the impactor. Assuming an isothermal atmosphere, which is a good approximation for the current Earth, the atmospheric mass inside a cone defined by angle $\theta$ measured from the normal of the impact site (see Figure \ref{fig14}) is given by
\begin{equation}\label{sp1}
\mathcal{M}_{Eject, \theta}=2 \pi \rho_0 \int^{a=\infty}_{a=0} \int^{\theta'=\theta}_{\theta'=0} \exp[-z/h] \sin \theta' a^2 d\theta' da
\end{equation}
where $\rho_0$ is the atmospheric density at the surface of the planet and $z$ is the height in the atmosphere above the ground and is related to $a$, the distance from the impact site to the top of the atmosphere (see Figure \ref{fig14}), by $z=(a^2+2aR\cos\theta')/2R$. Integrating over the whole cap, i.e. from $\theta=0$ to $\theta=\pi/2$, yields a total cap mass of
\begin{equation}
\mathcal{M}_{cap} = 2 \pi \rho_0 h^2 R,
\end{equation}
in the limit that $R \gg h$, which applies for the terrestrial planets. This is the maximum atmospheric mass that a single planetesimal impact can eject and is given by all the mass above the tangent plane of the impact site. The ratio of the mass in the cap compared to the total atmospheric mass is therefore $\mathcal{M}_{cap}/M_{atmos}=h/2R$. Atmospheric loss is therefore limited to at best $h/2R$ of the total atmosphere.

For impact velocities comparable to the escape velocity, the impactor mass needed to eject all the mass in the section of the cap subtended by $\theta$ is
\begin{equation}\label{e101}
m_{Imp,\theta}=2 \pi \rho_0 \int^{\infty}_{0} \exp[-(a^2+2aR\cos\theta)/2Rh]  a^2 da.
\end{equation}
Note, the integration in Equation (\ref{e101}) is only over $a$ and not $\theta$ since the explosion at the impact site is assumed to be isotropic (see Figure \ref{fig14}). Therefore the impactor mass needed to eject all the atmospheric mass above the tangent plane, $m_{Imp,\pi/2}=m_{cap}$, is
\begin{equation}\label{bv1xx}
\frac{m_{cap}}{\mathcal{M}_{cap}}= \left( \frac{\pi R}{2 h} \right)^{1/2},
\end{equation}
where we again assume that $R \gg h$. The impactor mass needed to eject all the mass above the tangent plane is about $\sqrt{R/h}$ larger than the mass in the cap. This is because the atmospheric mass close to the tangent plane is harder to eject due to its higher column density. Hence, in order to eject the entire cap an impactor of mass $m_{cap}=(\pi h/8R)^{1/2} M_{atmos}$ is needed. Evaluating this for the current Earth yields $m_{cap}= \sqrt{2}\rho_0 (\pi h R)^{3/2} \sim 3 \times 10^{-8} M_{\earth}$, which corresponds to impactor radii of $r_{cap}= (3 \sqrt{2 \pi} \rho_0/4 \rho)^{1/3} (hR)^{1/2}\sim 25~\rm{km}$ for impactor bulk densities of $\rho=2~\rm{g/cm^2}$.

\begin{figure}[htp]
\centering
   \includegraphics[width=0.7\textwidth]{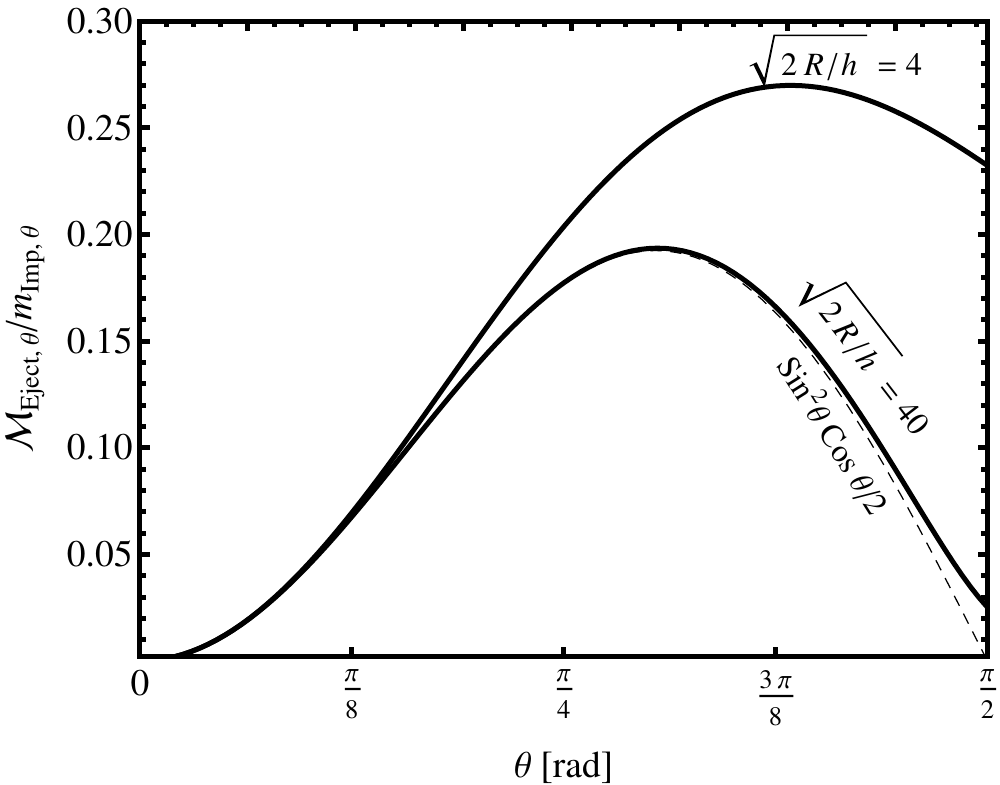}
\caption{Ratio of ejected mass, $\mathcal{M}_{Eject, \theta}$, to impactor mass, $m_{Imp, \theta}$, as a function of $\theta$. The solid lines correspond to an Earth-like planet, i.e. $\sqrt{2R/h}=40$, and an example of a close-in exoplanet with a scale height that is about 10\% of its radius,$\sqrt{2R/h}=4$. Close to the tangent plane (i.e., large $\theta$) larger impactor masses are needed because of the higher atmospheric column densities close to the tangent plane. The dashed line gives the analytic limit for $\theta \ll \pi/2-\sqrt{h/R}$.}
\label{fig103}
\end{figure}

Integrating and evaluating Equation (\ref{e101}) for $\theta=0$, yields $m_{min}=m_{imp,0}=4 \pi \rho_0 h^3$. For the current Earth this evaluates to $r_{min}=(3 \rho_0/\rho)^{1/3} h \sim 1~\rm{km}$. Impactors have to be larger than $r_{min}$ to be able to eject any atmosphere. For $\theta$ not too close to $\pi/2$, specifically  $\pi/2-\theta \gg \sqrt{h/R}$ (i.e., for $r/r_{min} \ll \sqrt{R/h}$), the ratio between the ejected mass and the impactor mass is given by
\begin{equation}
\frac{\mathcal{M}_{Eject,\theta}}{m_{Imp,\theta}}=\frac{\sin^2 \theta \cos \theta}{2}
\end{equation}
and is shown in Figure \ref{fig103}. $\mathcal{M}_{Eject,\theta}/m_{Imp,\theta}$ has a maximum at intermediate values of $\theta$, this is because for small $\theta$ the ejection efficiency is low because only a small fraction of the isotropic shock at the impact site is in the direction of $\theta$ for which the atmosphere can be ejected. In addition, for large $\theta$ the ejection efficiency is also low because significantly larger impactors are needed to eject the atmospheric mass along the tangent plane of the impact site due to its higher atmospheric column density. For small $\theta$, $\mathcal{M}_{Eject,\theta}/m_{Imp,\theta}$ can be approximated as
\begin{equation}\label{ex2}
\frac{\mathcal{M}_{Eject,\theta}}{m_{Imp,\theta}} \simeq \frac{r_{min}}{2r} \left(1-\left(\frac{r_{min}}{r}\right)^2 \right).
\end{equation}

\begin{figure}[htp]
\centering
   \includegraphics[width=0.8\textwidth]{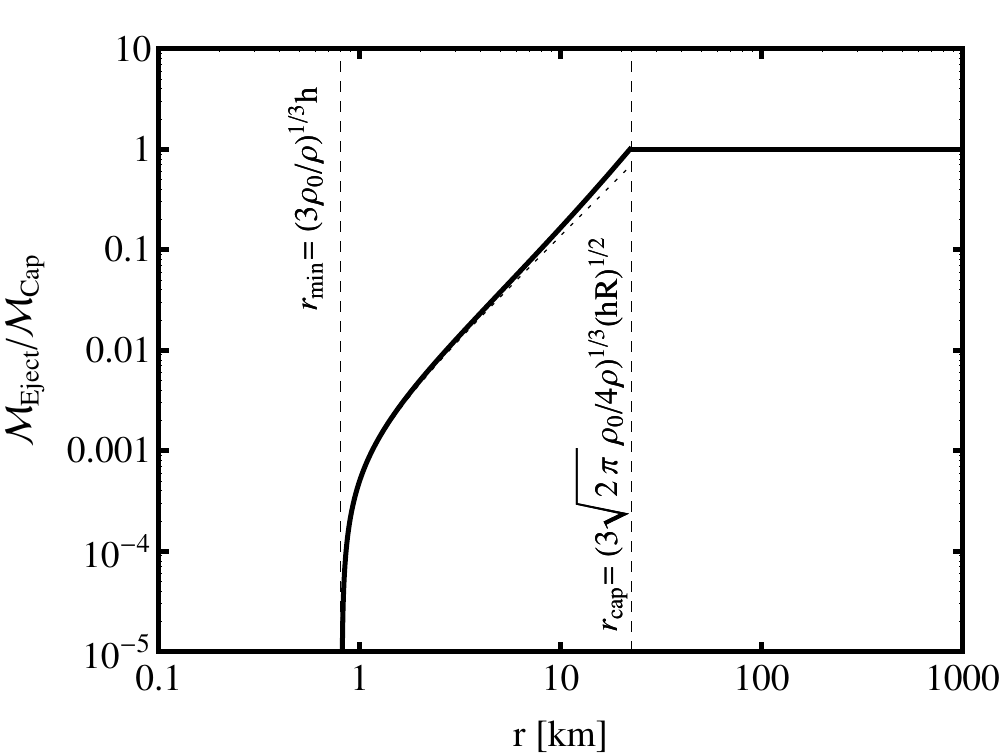}
\caption{Mass ejected in a single impact, $\mathcal{M}_{Eject}$, as a function of impactor radius, $r$. Only impactors with $r \geq r_{cap}$ are able to eject the whole cap. For the Earth this corresponds to impactors with $r \gtrsim 25~\rm{km}$. Impactors with $r_{min} < r < r_{cap}$ only eject a fraction of the atmospheric mass above the tangent plane of the impact site. For the Earth this corresponds to impactors with 1~km$ < r <$ 25~km.  Impactors smaller than $r_{min}$ (i.e., $r \lesssim 1~\rm{km}$) cannot eject any atmosphere. The dotted line that is close to the solid black curve corresponds to the small impactor limit derived in Equation (\ref{ex2}).}
\label{fig10}
\end{figure}

In summary, atmospheric erosion due to planetesimals therefore occurs in two different regimes. In the first regime, which was previously studied by \citet{MV89}, the planetesimals have masses large enough such that they can eject all the atmosphere above the tangent plane, in this case the planetesimal masses must satisfy $m \geq m_{cap}=\sqrt{2}\rho_0 (\pi h R)^{3/2}$. In the second regime, planetesimal impacts can only eject a fraction of the atmosphere above the tangent plane and their masses must satisfy $4 \pi \rho_0 h^3 < m < \sqrt{2}\rho_0 (\pi h R)^{3/2}$. As we discuss in section 5 and show in Figure \ref{fig13}, these small planetesimals are the most efficient impactors (per unit mass) for removing planetary atmospheres and may actually dominate the mass loss. Planetesimals with masses less than $m_{min}=m_{imp,0}=4 \pi \rho_0 h^3$ do not contribute to the atmospheric mass loss. Figure \ref{fig10} shows the atmospheric mass that can be ejected in a single planetesimal impact as a function of planetesimal size.

Our simple planetesimal impact model assumes an isotropic expansion of the vapor from the impact site. However, numerical simulations of planetesimal impacts show a strong preference for vertical expansion velocities \citep[e.g.][]{Shu09} and find significantly lower atmospheric mass loss for vertical impacts \citep{S07} compared to oblique ones  \citep{Shu09}. In contrast, in oblique impacts, the plume expands more isotropically and hence accelerates and ejects more atmospheric mass \citep{Shu09}. Comparing the results of our simple planetesimal impact model with the numerical results, averaged over all impact angles, obtained by \citet{Shu09}\footnote{The dimensionless erosional efficiency given in Equation (2) of \citet{Shu09} seems to contain a typo, since in its printed form it is not dimensionless. When comparing our results with \citet{Shu09} we assume that the author intended to have $\rho^2$ in denominator rather than just $\rho$, where $\rho$ is the density of the impactor.}, we find that we overestimate $\mathcal{M}_{Eject}/m_{Imp}$ by a factor of 10, 3 and 1 for impact velocities of 15~km/s, 20~km/s and 30~km/s, respectively. In deriving Equation (\ref{e101}), we assume that impact velocities comparable to $v_{esc}$ are sufficient to result in a point like explosion, where a mass equal to the mass of the impactor propagates isotropically with velocity of order $v_{esc}$, but comparison with numerical impact simulations above suggests that impactor velocities of about $3v_{esc}$ are needed to produce such an explosion. We did not investigate the dependence of $\mathcal{M}_{Eject}/m_{Imp}$ on the impact velocity. Previous works of numerical impact simulations find that bigger impact velocities lead to larger atmospheric mass loss, smaller values for $r_{min}$ and $r_*$ \citep{S07,Shu09}. From Equation (\ref{ex2}) we find that $\mathcal{M}_{Eject}/m_{Imp}$ has a maximum at $r_*=\sqrt 3 r_{min}$, which corresponds to about 2~km for the current Earth. This compares well the values of $r_*$ found by \citet{Shu09} which are 2~km, 1~km, and 1~km for impact velocities of 15~km/s, 20~km/s and 30~km/s, respectively. Finally, the scaling of $\mathcal{M}_{Eject}/m_{Imp}$ shown in Figure 3 of \citet{Shu09} is consistent with the $\mathcal{M}_{Eject}/m_{Imp}\propto m_{Imp}^{-1/3}$ scaling we find from Equation (\ref{ex2}) for $r_*<r<r_{cap}$ and the $\mathcal{M}_{Eject}/m_{Imp}\propto m_{Imp}^{-1}$ scaling we find for $r_{cap}<r$ (see also Figure \ref{fig13}).

\subsection{Impactor Size Distributions}
Similar to \citet{MV89}, we can now calculate the atmospheric mass loss rate due to planetesimal impacts for a given impactor flux. Parameterizing the cumulative impactor flux with a single power law given by $N(>r)=N_0 (r/r_0)^{-q+1}$, where $q$ is the differential power law index, $N_0$ is the impactor flux (number per unit time per unit area) normalized to impactors with radii $r_0$, we can write the atmospheric mass loss rate as
\begin{equation}\label{ex3}
\frac{dM_{atmos}}{dt}=- \pi R^2 \frac{N_0 (q-1)}{r_0} \int^{r_{max}}_{r_{min}} \left(\frac{r}{r_0}\right)^{-q} \mathcal{M}_{Eject}(r) dr.
\end{equation}
If the planetesimal size distribution is dominated by the smallest bodies such that $q>3$ then 
\begin{equation}
\frac{dM_{atmos}}{dt}=- \pi R^2 \frac{N_0 (q-1)m_{min}}{2 r_0}  \int^{r_{max}}_{r_{min}} \left(\frac{r}{r_0}\right)^{-q} \left( \left( \frac{r}{r_{min}} \right)^2-1 \right) dr
\end{equation}
where we substituted for $\mathcal{M}_{eject}$ from Equation (\ref{ex2}). Integrating over $r$ gives
\begin{equation}\label{ex1}
\frac{dM_{atmos}}{dt}=- \pi R^2 \frac{N_0 m_{min}}{q-3}  \left(\frac{r_{min}}{r_0}\right)^{-q+1}
\end{equation}
where $r_{min}=(3 \rho_0/\rho)^{1/3} h$ and $m_{min}=4 \pi \rho_0  h^3$. Evaluating Equation (\ref{ex1}) for $q=4$ yields $ dM_{atmos}/dt = -\pi R^2 N_0 \frac{4\pi}{3} \rho r_0^3$.

If $q<3$ then the atmospheric mass loss is dominated by impactors whose mass is around the smallest mass that can eject the entire cap. For this case we find for $3 > q > 1$
\begin{equation}\label{ex4}
\frac{dM_{atmos}}{dt}= - \pi R^2 C N_0 \mathcal{M}_{cap} \left(\frac{r_{cap}}{r_0}\right)^{-q+1},
\end{equation}
where $r_{cap}=(3\sqrt{2 \pi}\rho_0/4 \rho)^{1/3} (hR)^{1/2}$ is the impactor radius that can eject all the atmosphere above the tangent plane and $\mathcal{M}_{cap}=2 \pi \rho_0 h^2 R$ is the mass of the atmosphere above the tangent plane. $C$ is a constant that accounts for the additional contribution to the atmospheric mass loss from bodies that can only eject a fraction of the atmosphere above the tangent plane. $C=1$ implies that bodies smaller than $r_{cap}$ do not contribute to the atmospheric mass loss for $3 > q >1$. The numerical value of $C$ depends on the impactor size distribution because it is the bodies that are just a little bit smaller than $r_{cap}$ that can still contribute significantly to the atmospheric mass loss. We find that the values for $C$ range from 2.8 for $q=2.8$, 1.9 for $q=2.5$, 1.3 for $q=2.0$, to 1.1 for $q=1.5$. As expected, the value of $C$ is largest for $q$ close to 3 because the larger $q$, the more numerous are the smaller bodies.

The time it takes to lose the entire atmosphere is finite, i.e. the mass in the atmosphere does not simply decline exponentially towards zero but reaches zero in a finite time \citep{MV89}. This is because as some of the atmosphere is lost, its density declines and even smaller impactors can now contribute to the atmospheric mass loss. This accelerates the mass loss process, because smaller impactors are more numerous and dominate the mass loss (see Equations (\ref{ex1}) and (\ref{ex4})). From Equations (\ref{ex1}) and (\ref{ex4}) we find that for both $q>3$ and $1<q<3$ impactor size distributions that the rate of atmospheric mass loss scales as $M_{atmos}/dt \propto -M_{atmos}^{(-q+4)/3}$ and has a solution given by 
\begin{equation}\label{ed1}
M_{atmos}(t)=M_0 \left( 1-\frac{t}{t^*} \right)^{3/(q-1)},
\end{equation}
where $M_0$ is the initial atmospheric mass at $t=0$ and $t_*$ is the time it takes to lose the entire atmosphere.
Interestingly the solutions to Equation (\ref{ed1}) for both $q>3$ and $1<q<3$ only differ by the value of $t_*$. For $1<q<3$
\begin{equation}\label{exw1}
t^*_{q<3}=\frac{6}{ \pi (q-1) C Rh N_0}\left( \sqrt{\frac{\pi h}{8R}} \frac{M_0}{m_0} \right)^{(q-1)/3}
\end{equation}
and for $q>3$ the time for complete atmospheric loss is
\begin{equation}\label{exw2}
t^*_{q>3}=\frac{3 (q-3)}{ \pi (q-1) h^2 N_0}\left( \left(\frac{h}{R}\right)^2 \frac{M_0}{m_0} \right)^{(q-1)/3},
\end{equation}
where $m_0=4\pi \rho r_0^3/3$ and $r_0$ is the radius to which the size distribution is normalized. The expression in Equation (\ref{exw1}) differs from the one derived by \citet{MV89} because they assumed $\mathcal{M}_{cap} = m_{cap}$, whereas we find that $\mathcal{M}_{cap} = m_{cap} (2 h/\pi R)^{1/2}$ (see Equation (\ref{bv1xx})), and they neglected the numerical coefficient $C$.

\section{Comparison of Atmospheric Mass Loss due to Giant Impacts and Planetesimal Accretion}
Having derived the atmospheric mass loss due to giant impacts and smaller planetesimal impacts, we are now in the position to compare these different mass loss regimes. 

Assuming that all impactors have the same size, we find for $r_{min} < r < r_{cap}$ that the number of impactors needed to remove the atmosphere is
\begin{equation}\label{eh1}
N=\frac{M_{atmos}}{\mathcal{M}_{Eject}}=6 \frac{\rho_0 h}{\rho r_{min}}\left( \frac{R}{r}\right)^2 \left(1-\left(\frac{r_{min}}{r}\right)^2 \right)^{-1}
\end{equation}
and that this corresponds to a total mass in impactors given by
\begin{equation}\label{eh2}
M_T=\frac{M_{atmos}m_{Imp}}{\mathcal{M}_{Eject}}=  \frac{2r}{r_{min}} \left(1-\left(\frac{r_{min}}{r}\right)^2 \right)^{-1} M_{atmos}.
\end{equation}
Strictly speaking the Equations (\ref{eh1}) and (\ref{eh2}) overestimate $N$ and $M_T$, because as a fraction of the remaining atmosphere is removed a given sized impactor is able to eject a larger fraction of the atmosphere above the tangent plane. In deriving Equations (\ref{eh1}) and (\ref{eh2}) we used Equation (\ref{ex2}) for the relationship between the ejected mass and the impactor mass, which is only valid for $r/r_{min} \ll \sqrt{R/h}$. Equations (\ref{eh1}) and (\ref{eh2}) are therefore not accurate for $r \sim r_{cap}$ but should still give a reasonable estimate for Earth-like atmospheres since the deviation between the approximation and full solution is small and only occurs in the vicinity around $r \sim r_{cap}$ (see Figure \ref{fig10}).

Similarly, for impactors large enough to remove the entire cap but not too large to be in the giant impact regime (i.e., $r_{cap} < r < r_{gi}$), we have
\begin{equation}\label{eh3}
N=\frac{M_{atmos}}{\mathcal{M}_{Eject}}=\frac{2R}{h}
\end{equation}
and
\begin{equation}\label{eh4}
M_T=\frac{M_{atmos}m_{Imp}}{\mathcal{M}_{Eject}}= \frac{4 \pi}{3} \rho r^3 \frac{2R}{h}.
\end{equation}
In contrast to the previous regime, $r_{min} < r < r_{cap}$, impactors with $r_{cap} < r < r_{gi}$ are always limited to ejecting the whole cap, so an impactor of a given size cannot eject more atmosphere as the total atmospheric mass declines with time. 

We estimate the impactor radius at which giant impacts are more efficient than smaller impacts in ejecting the atmosphere, by equating the atmospheric mass loss due to giant impacts to the atmospheric cap mass. Assuming that $v_{Imp} \sim v_{esc}$, we find by equating Equation (\ref{gi0}) to the fraction of the atmosphere above the tangent plane that $r_{gi} \simeq (2hR^2)^{1/3}$, which corresponds to impactors with radii of about 900~km for the current Earth. Finally, from Equation (\ref{gi0}) we have that in the giant impact regime (i.e., $r>r_{gi}$) 
\begin{equation}\label{gi1}
N=\frac{M_{atmos}}{\mathcal{M}_{Eject}}=X_{loss}^{-1} \simeq \frac{3R^3}{r^3}
\end{equation}
and
\begin{equation}\label{gi2}
M_T=\frac{M_{atmos}m_{Imp}}{\mathcal{M}_{Eject}} \simeq 4M = \rm{constant}.
\end{equation}
Equations (\ref{gi1}) and (\ref{gi2}) were derived in the limit that $X_{loss} \ll 1$ in a single giant impact.

\begin{figure}[htp]
\centering
   \includegraphics[width=0.8\textwidth]{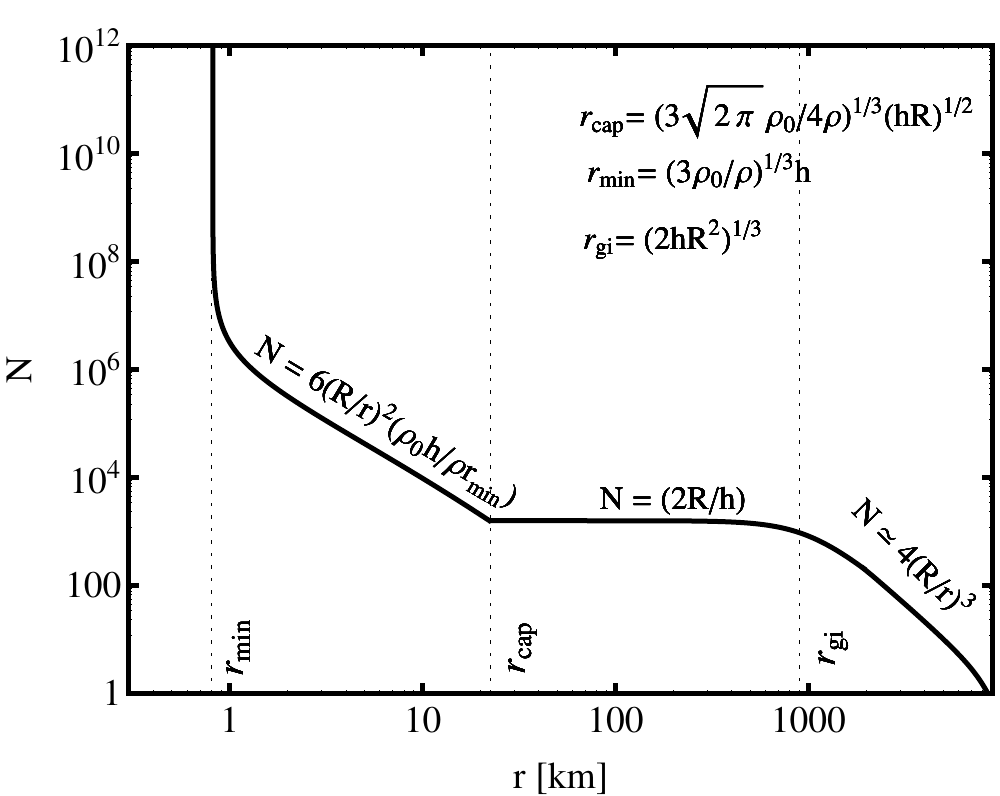}
\caption{Number of impactors needed, $N$, as a function of impactor radius, $r$, to eject the atmosphere, scaled to values of the current Earth.  Three distinct ejection regimes are apparent: 1) For small $r_{min} <r < r_{cap}$ (i.e., 1~km $\lesssim r \lesssim$ 25~km), the number of bodies needed scales roughly as $r^{-2}$. 2) For intermediate impactor sizes (i.e. 25~km$  < r < $1000~km), N is constant, because each impact ejects the whole atmospheric cap, and to eject the entire atmosphere one needs $N=M_{atoms}/\mathcal{M}_{cap}=(2R/h)$ number of impacts. 3) For larger impactor radii (i.e., $r>1000~\rm{km}$) the impactors are large enough to initiate a shock wave traveling through the entire Earth and launching a shock into the atmosphere globally such that $N$ tends to 1 as   $r$ tends to $R_{Earth}$. In the giant impact regime, $N \sim(R/r)^3$. Impactors with $r < r_{min} \sim 1~\rm{km}$ are not able to eject any atmosphere.}
\label{fig12}
\end{figure}

\begin{figure}[htp]
\centering
   \includegraphics[width=0.8\textwidth]{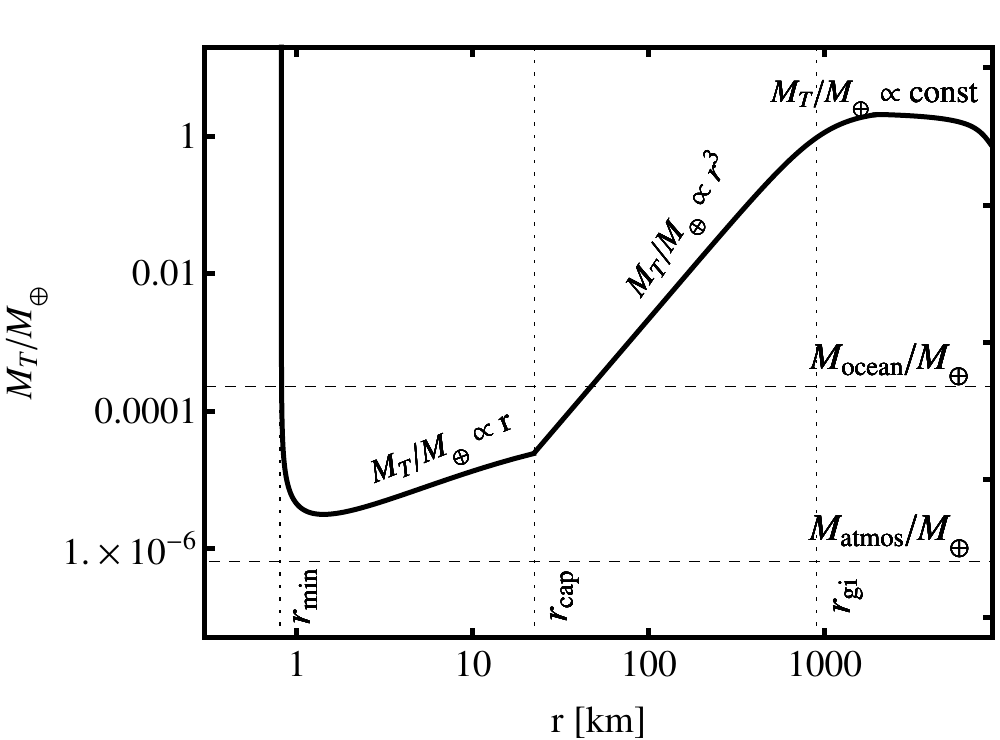}
\caption{Total impactor mass, $M_T$, needed to eject the atmosphere as a function of impactor radius, $r$. Several distinct ejection regimes are apparent, see caption of Figure \ref{fig12} for details. For comparison, the upper, middle, and lower dashed lines correspond to the mass ratio of the late veneer to the Earth's mass, the Earth's oceans to its total mass, and the Earth's atmosphere to its total mass, respectively. Small impactors with $r_* = \sqrt{3} r_{min}$ are the most efficient impactors per unit mass in ejecting the atmosphere (see Equation (\ref{ex2})). For the current Earth this corresponds to bodies with $r \sim 2~\rm{km}$. The ratio between the impactor mass to the atmospheric mass ejected for $r=r_*$ is $m_{Imp}/\mathcal{M}_{Eject} = 3^{3/2} \simeq 5$ (see Equation (\ref{ex2})). This implies that a planetesimal population comprised of bodies with $r \sim r_*$ would only need to contain about $5 M_{atmos}$ in mass to eject the planetary atmosphere. This is an absolute tiny amount compared to estimates of the mass in planetesimals during and even at the end of the giant impact phase of terrestrial planet formation. Impactors with $r< r_{min} \sim 1~\rm{km}$ are not able to eject any atmosphere. }
\label{fig11}
\end{figure}

\begin{figure}[htp]
\centering
   \includegraphics[width=0.8\textwidth]{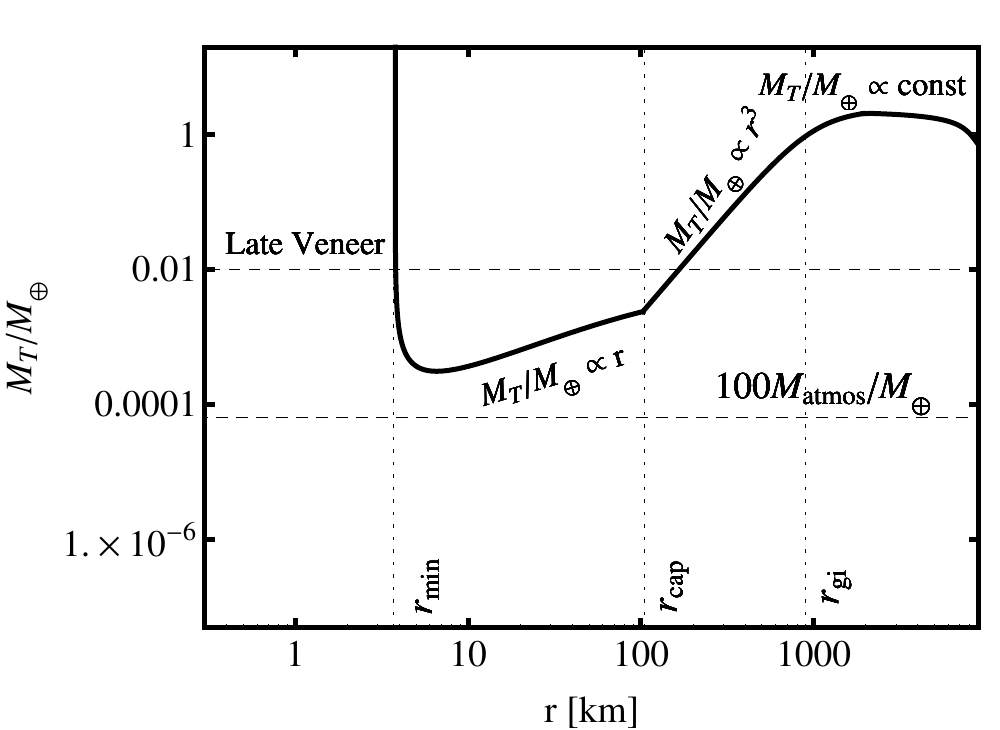}
\caption{Same as in Figure \ref{fig11} but for an atmospheric mass that is 100 times enhanced compared to that of the current Earth. For comparison, the upper and lower dashed lines correspond to the mass ratio of the late veneer to the Earth's mass and 100 times the Earth's current atmosphere to its total mass, respectively.}
\label{fig111}
\end{figure}

Figure \ref{fig12} shows the number of impactors needed, defined here as $N=M_{atoms}/\mathcal{M}_{Eject}$, to erode the atmosphere as a function of impactor radius. Figure \ref{fig11} shows the total mass in impactors needed, defined here as $M_T=M_{atoms}m_{Imp}/\mathcal{M}_{Eject}$, to erode the atmosphere as a function of impactor radius. Figure \ref{fig111} is the same as Figure \ref{fig11} but for atmospheric mass that is 100 times enhanced compared to that of the current Earth. The plots in all three figures assume that all impactors are identical and have a single size, $r$. Figures \ref{fig12}, \ref{fig11} and \ref{fig111} clearly display the three distinct ejection regimes. Figures \ref{fig11} and \ref{fig111} impressively show that small impactors with $r_{min}<r<r_{cap}$ are the most effective impactors per unit mass in ejecting the atmosphere. The best impactor size for atmospheric mass loss is $r_* = \sqrt{3} r_{min}$  for which $m_{Imp}/\mathcal{M}_{Eject} = 3^{3/2}  \simeq 5$. For the current Earth this corresponds to bodies with $r \sim 2~\rm{km}$ and implies that a total mass in such impactors only needs to be about $5 M_{atoms}$ to eject the planetary atmosphere. This is an absolutely tiny amount compared to estimates of the mass in planetesimals during and even at the end of the giant impact phase. The implications of our findings for terrestrial planet formation are discussed in section \ref{s5}.

\section{Application \& Importance for the Formation of the Terrestrial Planets}\label{s5}
Earth, Venus and Mars all display similar geochemical abundance patterns of near chondritic light noble gasses, but relative depletion of in $\rm{Xe}$, C and N \citep[e.g.][]{H13}. This suggests that all three planets may not only have lost major volatiles, but also accreted similar veneers from chondritic material. In addition, all three planets have similar noble gas patterns, but whereas the budgets for Venus are near chondritic, the budgets for Earth and Mars are depleted by two and four orders of magnitude, respectively. This suggests that Earth and Mars lost the vast majority of their noble gasses relative to Venus during the process of planet formation \citep{H13}.

Recent work suggests that the Earth went through at least two separate periods during which its atmosphere was lost \citep{TM14}. The evidence for several atmospheric loss events is inferred from the mantle $^3\rm{He}/^{22}\rm{Ne}$, which is higher than the primordial solar abundance by at least a factor of 6 and which is thought to have been increased to its current value by multiple magma ocean degassing episodes and atmospheric loss events. In addition, \citet{TM14} suggest that the preservation of low $^3\rm{He}/^{22}\rm{Ne}$ ratio in a primitive reservoir sampled by plumes implies that later giant impacts did not generate a global magma ocean.

Previous works usually appeal to giant impacts to explain Earth's atmospheric mass loss episodes \citep[e.g.][]{GA03,GA05}. Figure \ref{fig11}, however, demonstrates clearly that small planetesimals with sizes $r_{min} < r <  r_{cap}$ are the most efficient impactors per unit mass in ejecting the atmosphere. For the current Earth this corresponds to bodies with $1\rm{km}  \lesssim r \lesssim 25~\rm{km}$. Furthermore, atmospheric mass loss due to small impactors will proceed without generating a global magma ocean, which is supported by recent interpretations of low $^3\rm{He}/^{22}\rm{Ne}$ ratios in a primitive reservoir sampled by plumes \citep{TM14}.

\begin{figure}[htp]
\centering
   \includegraphics[width=0.8\textwidth]{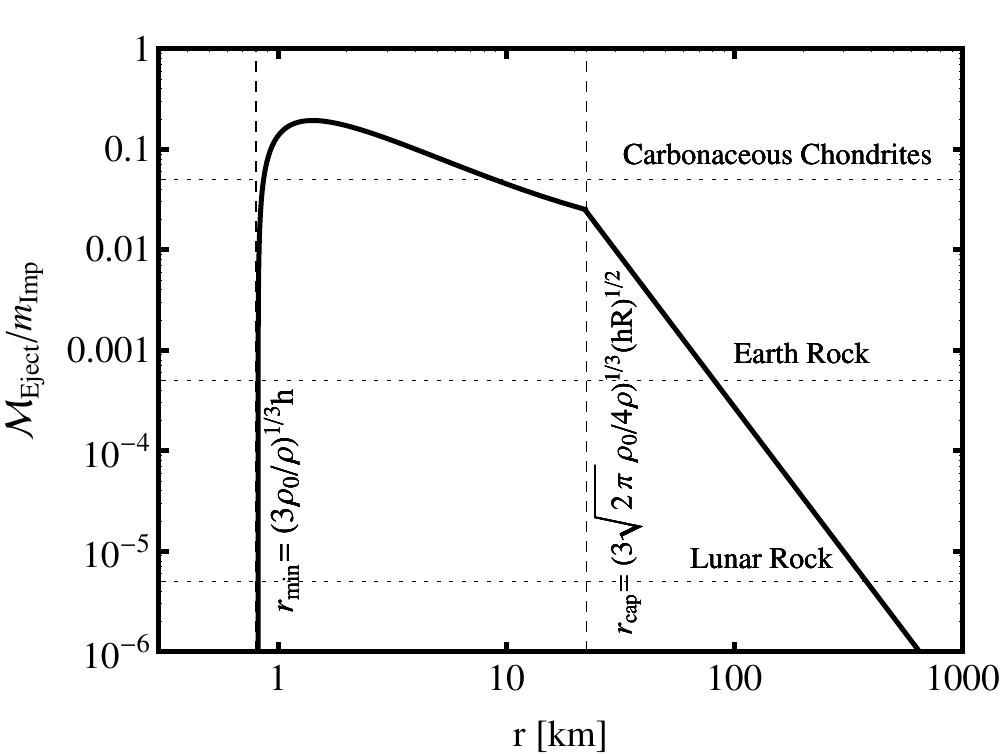}
\caption{Ratio of atmospheric mass ejected to impactor mass, $\mathcal{M}_{Eject}/m_{Imp}$. Numerical values are scaled to the current Earth. Small impactors with $r_* = \sqrt{3} r_{min}$ are the most efficient impactors per unit mass in ejecting the atmosphere (see Equation (\ref{ex2})). For the current Earth this corresponds to bodies with $r \sim 2~\rm{km}$. The ratio between the impactor mass to the atmospheric mass ejected for $r=r_*$ is $m_{Imp}/\mathcal{M}_{Eject} = 3^{3/2} \simeq 5$ (see Equation (\ref{ex2})). The value of $\mathcal{M}_{Eject}/m_{Imp}$ decreases rapidly for larger planetesimals. Whether or not planetesimal impacts will lead to a net loss of planetary atmospheres depends on the impactor sizes distribution as well as their volatile budget. The three dotted horizontal lines correspond to volatile contents of 5 wt.\% (representative of some of the most water rich carbonaceous chondrites), 0.05 wt.\% (representative of the average water content in the bulk Earth excluding the hydrosphere) and 0.0005 wt.\%  corresponding to an estimate of the minimum water content of the bulk moon \citep{MCS10}.}
\label{fig13}
\end{figure}

Whether or not planetesimal impacts will lead to a net loss of planetary atmospheres or simply an alteration of the current atmosphere depends on the planetesimal sizes distribution as well as the volatile content of the planetesimals. \citet{ZP92} investigated impact erosion and replenishment of planetary atmospheres and suggest that the competition of these two processes can explain the present distributions of atmospheres between Ganymede, Callisto, and Titan. \citet{NK12} performed a similar study with a focus on Earth and Mars during a heavy bombardment and find a dominance of accumulation over erosion. Figure \ref{fig13} shows the ratio of atmospheric mass ejected to impactor mass as a function of planetesimal size. 
If the impactors are not dominated by a single size, as assumed in Figure \ref{fig13}, but instead follow a power-law size distribution, $N(>r)=N_0 (r/r_0)^{-q+1}$, then the ratio of the atmospheric mass lost to the impactor mass is, for $3<q<4$, given by
\begin{equation}\label{ex7}
\frac{dM_{atmos}}{dm_{Imp}}= -\frac{4-q}{(q-1)(q-3)} \left( \frac{r_{min}}{r_{max}}\right)^{-q+4}+f,
\end{equation}
where $r_{max}$ is the maximum size of the planetesimal size distribution and $r_{min}=(3 \rho_0/\rho)^{1/3}h$ is the smallest planetesimal size that can contribute to the atmospheric mass loss as derived in section \ref{s1} and $f$ is the volatile fraction of the planetesimals. Similarly, for $1<q<3$ we have
\begin{equation}\label{ex8}
\frac{dM_{atmos}}{dm_{Imp}}= -C \left( \frac{2h}{\pi R}\right)^{1/2}\frac{4-q}{q-1} \left( \frac{r_{cap}}{r_{max}}\right)^{-q+4} +f,
\end{equation}
where $r_{cap}= (3 \sqrt{2 \pi} \rho_0/4 \rho)^{1/3} (hR)^{1/2}$ and corresponds to the impactor radius that can eject all the atmospheric mass above the tangent plane. Evaluating the first term in Equations (\ref{ex7}) and (\ref{ex8}) for a planetesimal population ranging from $r<r_{min} \sim 1~\rm{km}$ to 1000~km and assuming values of the current Earth we find $dM_{atmos}/dm_{Imp} =- 0.01+f$  for $q=3.5$ and  $dM_{atmos}/dm_{Imp} = -0.0003+f$ for $q=2.5$, respectively \footnote{For comparison, the lunar craters can be modeled with a power-law size distribution with $q\sim 2.8$ and $q\sim 3.2$ for crater diameters ranging from 1~km to 64~km and larger than 64~km, respectively \citep[e.g.][]{N01}.}. These results have two important implications: First, we can estimate how massive initial planetary atmospheres must have been in order to avoid erosion due to planetesimal impacts. Estimates of the mass in planetesimals during the giant impact phase range from a few percent to several tens of percent of the total mass in terrestrial planets \citep[e.g.][]{SWY12}. Assuming a total mass in planetesimals of about $0.1~M_{\earth}$ yields that initial atmospheres must have contained $M_{atmos} \gtrsim 10^{-3}M_{\earth}$ and  $M_{atmos} \gtrsim 3 \times 10^{-5}M_{\earth}$ for $q=3.5$ and $q=2.5$, respectively, in order to avoid erosion due to planetesimal impacts. The latter result is particular interesting since it implies that for $q=2.5$ Venus, which has $M_{atmos} \sim 8 \times 10^{-5} M_{\earth}$, will not undergo atmospheric erosion due to planetesimal impacts whereas the Earth could have lost most of its atmosphere due to planetesimal impacts if its initial atmosphere was less than $ 3 \times 10^{-5}M_{\earth}$. Second, Equations (\ref{ex7}) and (\ref{ex8}) permit an equilibrium solution, where the atmospheric erosion is balanced by the volatiles delivered to the planet's atmosphere in a given planetesimal impact. It may therefore be that the Earth's atmosphere was eroded by planetesimal impacts until an equilibrium was established between atmospheric loss and volatile gain. The current Earth's atmosphere could be the result of such an equilibrium if the fraction of the planetesimal mass that ends up as volatiles in the atmosphere, $f$, was 0.01 and $3 \times 10^{-4}$ for $q=3.5$ and $q=2.5$, respectively. These finding are consistent with results by \citet{NK12} who find that atmospheric erosion is balanced by volatile delivery from an asteroidal population of impactors if $f=2 \times 10^{-3}$.

To summarize, we have shown that planetesimals can be very efficient in atmospheric erosion and that the amount of atmospheric loss depends on the total mass in planetesimals, on their size distribution and their volatile content. The total  planetesimal mass needed for significant atmospheric loss is small and it is therefore likely that planetesimal impacts played a major role in atmospheric mass loss over the formation history of the terrestrial planets. We have shown that the current differences in Earth's and Venus' atmospheric masses can be explained by modest differences in their initial atmospheric masses and that the current atmosphere of the Earth could have resulted from an equilibrium between atmospheric erosion and volatile delivery to the atmosphere by planetesimal impacts. Furthermore, if the Earth's hydrosphere was dissolved in its atmosphere, as it may have been immediately after a giant impact, then planetesimal impacts can also have contributed significantly to loss of the Earth's oceans. We have shown above that planetesimals can be very efficient in atmospheric erosion and that the amount of atmospheric loss depends both on the total mass in planetesimals, on their size distribution and their volatile content. One way for planetesimals to not participate significantly in the atmospheric erosion of some, or all, of the terrestrial planets is for most of their mass to reside in bodies smaller than $r_{min}=(3\rho_0/\rho)^{1/3}h$, since such bodies are too small to contribute to atmospheric loss. Finally, planetesimal impacts may not only have played a major role in atmospheric erosion of the terrestrial planets but may also have contributed significantly to the current terrestrial planet atmospheres.

\section{Discussion \& Conclusions}
We investigated the atmospheric mass loss during planet formation and found that it can proceed in three different regimes. 

1) In the first regime ($r \gtrsim r_{gi}=(2h R^2)^{1/3}$), giant impacts create strong shocks that propagate through the planetary interior causing a global ground motion of the protoplanet. This ground motion in turn launches a strong shock into the planetary atmosphere, which can lead to loss of a significant fraction or even the entire atmosphere. We find that the local atmospheric mass loss fraction due to giant impacts for ground velocities $v_g \lesssim 0.25 v_{esc}$ is given by $\chi_{loss}=(\beta v_g/v_{esc})^p$ where $\beta$ and $p$ are constants equal to $\beta=1.71$, p=4.9 (isothermal atmosphere and an adiabatic index $\gamma=5/3$) and $\beta=2.11$, p=7.6 (adiabatic atmosphere with polytropic index $n=1.5$, adiabatic index $\gamma=5/3$). In addition, using a simple model of a spherical shock propagating through the target, we find that the global atmospheric mass loss fraction is well characterized by $X_{loss} \simeq 0.4x+1.2x^2-0.8x^3$ (isothermal) and  $X_{loss} \simeq 0.4x+1.8x^2-1.2x^3$ (adiabatic), where $x=(v_{Imp}m/v_{esc}M)$, independent of the precise value of the adiabatic index.

2) In the second regime ($r_{cap}= (3\sqrt{2\pi}\rho_0/4 \rho)^{1/3}(hR)^{1/2}\lesssim r \lesssim (2h R^2)^{1/3}=r_{gi}$), impactors cannot eject the atmosphere globally, but are large enough, i.e., $r>r_{cap}$, to eject all the atmosphere above the tangent plane of the impact site. A single impactor is therefore limited to ejecting $h/2R$ of the total atmosphere in a given impact. For the current Earth this corresponds to impactor sizes satisfying $25~\rm{km} \lesssim r \lesssim 900~\rm{km}$.

3) In the third regime ($r_{min}= (3\rho_0/\rho)^{1/3}h\lesssim r \lesssim (3\sqrt{2\pi}\rho_0/4 \rho)^{1/3}(hR)^{1/2}=r_{cap}$),  impactors are only able to eject a fraction of the atmospheric mass above the tangent plane of the impact site. For the current Earth this corresponds to 1~km $\lesssim r \lesssim$ 25~km. Impactors with $r \lesssim r_{min}$ are not able to eject any atmosphere. 

Comparing these three atmospheric mass loss regimes, we find that the most efficient impactors (per unit impactor mass) for atmospheric loss are small planetesimals. For the current atmosphere of the Earth this corresponds to impactor radii of about 2~km. For such impactors, the ejected mass to impactor mass ratio is only $\sim 5$, implying that one only needs about 5 times the total atmospheric mass in such small impactors to active complete loss. More realistically, planetesimal sizes were probably not constrained to a single size, but spanned by a range of sizes. For impactor flux size distributions parametrized by a power law, $N>r \propto r^{-q+1}$, with differential power law index $q$ we find that  for $1<q<3$ the atmospheric mass loss is dominated by bodies that eject all the atmosphere above the tangent plane ($r>r_{cap}$) and that for $q>3$ the mass loss is dominated by impactors that only erode a fraction of the atmospheric mass above the tangent plane in a single impact ($r_{min}< r<r_{cap}$). Assuming that the planetesimal population ranged in size from $r<r_{min} \sim 1~\rm{km}$ to 1000~km, we find for, parameters corresponding to the current Earth, an atmospheric mass loss rate to impactor mass rate ratio of 0.01 and 0.0003 for $q=3.5$ and $q=2.5$, respectively. Despite being bombarded by the same planetesimal population, we find that the current differences in Earth's and Venus' atmospheric masses can be explained by modest differences in their initial atmospheric masses and that the current atmosphere of the Earth could have resulted from an equilibrium between atmospheric erosion and volatile delivery to the atmosphere from planetesimal impacts. 
 
Recent work suggests that the Earth went through at least two separate periods during which its atmosphere was lost and that later giant impacts did not generate a global magma ocean \citep{TM14}. Such a scenario is challenging to explain if atmospheric mass loss was a byproduct of giant impacts, because a combination of large impactor masses and large impact velocities is needed to achieve complete atmospheric loss (see Figure \ref{fig7}). Furthermore, giant impacts that could accomplish complete atmospheric loss, almost certainly will generate a global magma ocean. Since atmospheric mass loss due to small planetesimal impacts will proceeded without generating a global magma ocean they offer a solution to this conundrum. 

To conclude, we have shown that planetesimals can be very efficient in atmospheric erosion and that the amount of atmospheric loss depends on the total mass in planetesimals, on their size distribution and their volatile content. The total  planetesimal mass needed for significant atmospheric loss is small and it is therefore likely that planetesimal impacts played a major role in the atmospheric mass loss history of the Earth and during planet formation in general. In addition, small planetesimal impacts may also have contributed significantly to the current terrestrial planet atmospheres.

{\bf Acknowledgements:} We thank H. J. Melosh and the second anonymous referee for their constructive reviews and D. Jewitt, T. Grove, N. Inamdar for helpful comments and suggestions. RS dedicates this paper to the late Tom Ahrens, who initiated his interest in the problem of atmospheric escape and collaborated on related ideas.

\bibliographystyle{aj} 
%\bibliography{spin} 

\end{document}